\documentclass[aps,reprint,prx,superscriptaddress]{revtex4-2}
\usepackage{amsmath}
\usepackage{amsfonts}
\usepackage{amssymb}
\usepackage{physics}
\usepackage{arydshln}
\usepackage{array}
\usepackage{subfiles}
\usepackage{hyperref}
\usepackage{graphicx}
\usepackage{caption}
\usepackage{subcaption}
\usepackage{tabularx}
\usepackage{booktabs}
\usepackage{color}
\graphicspath{{./figures/}}

\begin{document}
\title{Neural Wave Functions for Superfluids}
\author{Wan Tong Lou}
\author{Halvard Sutterud}
\author{Gino Cassella}
\author{W.M.C.~Foulkes}
\affiliation{
  Department of Physics,
  Imperial College London,
  South Kensington Campus,
  London SW7 2AZ,
  United Kingdom}
\author{Johannes Knolle}
\affiliation{
  Department of Physics,
  Imperial College London,
  South Kensington Campus,
  London SW7 2AZ,
  United Kingdom}
\affiliation{
  Department of Physics TQM,
  Technische Universit\"{a}t M\"{u}nchen,
  James-Franck-Stra{\ss}e 1,
  D-85748 Garching,
  Germany}
\affiliation{
  Munich Center for Quantum Science and Technology (MCQST),
  80799 Munich,
  Germany}
\author{David Pfau}
\affiliation{
  DeepMind, 
  6 Pancras Square, 
  London N1C 4AG, 
  United Kingdom}
\affiliation{
  Department of Physics,
  Imperial College London,
  South Kensington Campus,
  London SW7 2AZ,
  United Kingdom}
\author{James S.\ Spencer}
\affiliation{
  DeepMind, 
  6 Pancras Square, 
  London N1C 4AG, 
  United Kingdom}
\date{\today}

\begin{abstract}
Understanding superfluidity remains a major goal of condensed matter physics.
Here we tackle this challenge utilizing the recently developed Fermionic neural network (FermiNet) wave function Ansatz \cite{pfau2020} for variational Monte Carlo calculations.
We study the unitary Fermi gas, a system with strong, short-range, two-body interactions known to possess a superfluid ground state but difficult to describe quantitatively.
We demonstrate key limitations of the FermiNet Ansatz in studying the unitary Fermi gas and propose a simple modification based on the idea of an antisymmetric geminal power singlet (AGPs) wave function. The new AGPs FermiNet outperforms the original FermiNet significantly in paired systems, giving results which are more accurate than fixed-node diffusion Monte Carlo and are consistent with experiment.
We prove mathematically that the new Ansatz, which only differs from the original Ansatz by the method of antisymmetrization, is a strict generalization of the original FermiNet architecture, despite the use of fewer parameters.
Our approach shares several advantages with the original FermiNet: the use of a neural network removes the need for an underlying basis set; and the flexibility of the network yields extremely accurate results within a variational quantum Monte Carlo framework that provides access to unbiased estimates of arbitrary ground-state expectation values. We discuss how the method can be extended to study other superfluids.
\end{abstract}

\maketitle

\section{Introduction}
\label{section:introduction}

The unitary Fermi gas (UFG) is a paradigmatic example of a strongly interacting system of two-component fermions that possesses a superfluid ground state and lies in the crossover region between a Bardeen-Cooper-Schrieffer (BCS) superconductor and a Bose-Einstein condensate \cite{leggett1980,nozieres1985}.
The effective range of the interaction is zero and the $s$-wave scattering length diverges (the ``unitarity limit''), so the UFG has no intrinsic length scale.
The only remaining length is the inverse of the Fermi wavevector $1/k_F$, on which all thermodynamic quantities depend.
For example, regardless of the particle density, the ground-state energy per particle of a unitary Fermi gas can be written as
\begin{equation}
    E = \xi E_{FG} = \xi \frac{3}{5} \frac{\hbar^2 k_F^2}{2m} ,
    \label{eq:bertsch}
\end{equation}
where $E_{FG}$ is the energy per particle of a non-interacting Fermi gas of the same density.
The dimensionless constant $\xi$ is known as the Bertsch parameter \cite{bertsch1999}.

Because of the universality of the UFG model, it can be used to describe many real physical systems at different scales, such as the neutron matter in the inner crust of a neutron star \cite{gandolfi2015} or the quantum criticality of an s-wave atomic superfluid~\cite{nishida2006,nikolic2007renormalization}.
The size of the pairs in the UFG is comparable to the inter-particle spacing, which is also a feature of many high-$T_c$ superconductors \cite{randeria1990,randeria2010,strinati2018}.
As a result, the UFG has been studied extensively \cite{giorgini2008}.
Although the UFG is an idealized model, it can be accurately realized in the laboratory using ultracold atomic gases in which the interactions have been tuned by using an external magnetic field to drive the system across a Feshbach resonance \cite{chin2010}. 

The UFG has been studied for decades, but it remains difficult to calculate its ground-state properties accurately using analytic methods.
Mean-field treatments such as BCS theory \cite{bcs} give good results for systems with weak interactions, but there is no guarantee of success in the strongly interacting regime.
As a result, various quantum Monte Carlo (QMC) methods \cite{foulkes2001,carlson_review_2013} have been used to simulate the properties of the UFG to high accuracy at zero and finite temperature.
Methods used include variational Monte Carlo (VMC), fixed-node diffusion Monte Carlo (FN-DMC), fixed-node Green function Monte Carlo, auxiliary field Monte Carlo and diagrammatic Monte Carlo \cite{carlson2003,chang_quantum_2004,morris_ultracold_2010,chang2005,gezerlis_low-density_2010,forbes2011,he_superfluid_2020,song2021,goulko2016,jensen2020,pisani2022}.
However, a full quantitative description remains an open and challenging problem.

Recent advances in machine learning algorithms and the growing availability of inexpensive GPU-based computational resources have allowed neural-network-based approaches to permeate many areas of computational physics, including lattice \cite{troyer2017,choo2019,hibat-allah2020,luo2019} and continuum \cite{pfau2020,hermann2020,gerard2022,pescia2022} QMC simulations.
Here we employ a neural network Ansatz within a VMC approach to study the unitary Fermi gas.
The Ansatz we use, the Fermionic Neural Network (FermiNet) \cite{pfau2020}, gives very accurate results for atoms and molecules \cite{pfau2020,li2022,wilson2021,spencer2020} and has recently been applied to periodic solids and the homogeneous electron gas (HEG) with comparable success \cite{li2022b}.
In the case of the HEG, the variational optimization of the FermiNet Ansatz discovered the quantum phase transition between the Fermi liquid and Wigner crystal ground states without external guidance \cite{gcassella2022}.
In contrast, previous approaches required different Ans\"{a}tze to be used for the two different phases.
The FermiNet has not previously been applied to fermionic superfluids such as the UFG.

The paper is organized as follows. Section \ref{section:ferminet} describes the architecture of the FermiNet.
We find that the original FermiNet Ansatz is insufficient to capture the two-particle correlations of superfluids.
Although a FermiNet wave function with one determinant and a sufficiently large neural network is in principle able to represent any fermionic state \cite{pfau2020}, it is often advantageous to use a network of a fixed size and a small linear combination of FermiNet determinants.
In the case of the unitary Fermi gas, however, we find that the number of block-diagonal determinants required to describe the ground state accurately scales exponentially with the system size.
This is the first example in which the FermiNet has been seen to fail both quantitatively and qualitatively, and suggests that the FermiNet wave function may not be able to represent arbitrary fermionic wave functions in practice.
To remedy the problem, we utilize the neural-network part of the FermiNet architecture to build a different type of wave function based on the idea of an antisymmetric geminal power singlet wave function (AGPs) \cite{bouchaud_pair_1988,casula_geminal_2003,bajdich_generalized_2007,genovese_general_2020,luo2019}, which we discuss in detail in Section \ref{section:agp}.
This leads to substantial improvements, even though the neural-network part of the wave function remains unchanged.
The implementation of the AGPs wave function using the FermiNet, as well as its relation to the original block-diagonal multi-determinant FermiNet, are discussed in Section \ref{section:agps_ferminet}.
Our computational results are presented in Section \ref{section:results}, followed by a summary and discussion in Section \ref{section:discussion}. The Appendix includes detailed explanations and derivations of important formulae, as well as implementation and training details.

\section{FermiNet}
\label{section:ferminet}

The Fermionic Neural Network, or FermiNet \cite{pfau2020}, is a neural network that can be used to approximate the ground-state wave function of any system of interacting fermions.
The inputs to the network are the positions $\mathbf{r}_1,\mathbf{r}_2,\ldots,\mathbf{r}_N$ and spin coordinates ${\sigma}_1, {\sigma}_2, \ldots, {\sigma}_N$ of the $N$ particles, and the output is the value of the wave function $\Psi(\mathbf{r}_1,{\sigma}_1,\mathbf{r}_2,{\sigma}_2,\ldots,\mathbf{r}_N,{\sigma}_N)$ corresponding to those inputs.
The network is trained using the variational Monte Carlo (VMC) method \cite{foulkes2001}:
the weights and biases that define the network are varied at each training iteration to minimize the energy expectation value according to the variational principle.
If the network is flexible enough, the approximate wave function obtained after training may be very close to the true ground state.
The FermiNet provides a more general and accurate alternative to the conventional Slater-Jastrow (SJ) and Slater-Jastrow-backflow (SJB) Ans\"{a}tze that have been used in most VMC and FN-DMC calculations to date, and may improve VMC and FN-DMC results for strongly correlated systems.

In conventional SJ Ans\"atze, the antisymmetry of the $N$-electron wave function is represented using Slater determinants, which are antisymmetrized products of single-particle orbitals.
For simulations of solids, it is common to use one determinant only; for molecules, a linear combination of determinants is usually employed.
In both cases, the presence of determinants guarantees that the wave function has the correct exchange antisymmetry.
To improve the representation of electronic correlations, especially the correlations that chemists call ``dynamic'', the determinants are multiplied by a totally symmetric non-negative function of the electron coordinates known as a Jastrow factor.
This acts to decrease the value of the wave function as pairs of electrons approach each other, reducing the total Coulomb repulsion energy.

If the Hamiltonian is independent of spin and all the single-particle orbitals are eigenfunctions of total $S_z$, one can assign spins to the electrons and every Slater determinant can be factored into a product of spin-up and spin-down Slater determinants \cite{toulouse2007,foulkes2001}.
The wave function is no longer antisymmetric under the exchange of electrons of opposite spin, but expectation values of spin-independent operators are unaltered.
Including a spin-assigned Jastrow factor expressed in the form $e^{J}$, a one-determinant SJ Ansatz becomes:
\begin{align}
    &\Psi_\text{SJ}\left(\{\mathbf{r}^{\uparrow}\},\{\mathbf{r}^{\downarrow}\}\right) \notag \\
    &= e^{J(\{\mathbf{r}^{\uparrow}\},\{\mathbf{r}^{\downarrow}\})} \det[\phi^{\uparrow}_{i}(\mathbf{r}^\uparrow_j)] \det[\phi^{\downarrow}_{i}(\mathbf{r}^\downarrow_j)] ,
    \label{eq:block-det-sd}
\end{align}
where $\{\mathbf{r}^{\uparrow}\}$ and $\{\mathbf{r}^{\downarrow}\}$ are the sets of position coordinates of the $N^{\uparrow}$ electrons assigned to be spin up and the $N^{\downarrow}$ electrons assigned to be spin down, respectively.

One can improve the SJ Ansatz by transforming the electron coordinates as
\begin{align}
  \mathbf{r}_j^{\alpha} \rightarrow \mathbf{x}_j^{\alpha} 
  = \mathbf{r}_j^{\alpha} 
  + \sum_{\stackrel{\scriptstyle i=1}{\scriptstyle (i \ne j)}}^{N^{\alpha}} 
  & \eta_{\parallel}( r_{ij}^{\alpha\alpha} )( \mathbf{r}_i^{\alpha} - \mathbf{r}_j^{\alpha} )
  \notag \\ 
  &+ \sum_{i=1}^{N^{\bar{\alpha}}} \eta_{\nparallel}( r_{ij}^{\bar{\alpha}\alpha} )
  ( \mathbf{r}_i^{\bar{\alpha}} - \mathbf{r}_j^{\alpha} ),
\end{align}
where $\alpha$ and $\bar{\alpha}$ are the two possible spin components of an electron, $r_{ij}^{\beta\alpha} = |\mathbf{r}_i^{\beta} - \mathbf{r}_j^{\alpha}|$, and $\eta_{\parallel}(r)$ and $\eta_{\nparallel}(r)$ are parameterized functions of a single distance argument.
The coordinate-transformed SJ Ansatz is called a Slater-Jastrow-backflow (SJB) wave function, and the new coordinates are called \emph{quasiparticle coordinates}.
Note that the quasiparticle coordinate $\mathbf{x}_j^{\alpha}$ is invariant under the exchange of any two position vectors in $\{\mathbf{r}_{/j}^{\alpha}\} = \{\mathbf{r}_{1}^{\alpha}, \dots, \mathbf{r}_{j-1}^{\alpha}, \mathbf{r}_{j+1}^{\alpha}, \dots, \mathbf{r}_{{N}^{\alpha}}^{\alpha}\}$ or in $\{\mathbf{r}^{\bar{\alpha}}\} = \{ \mathbf{r}_1^{\bar{\alpha}}, \dots, \mathbf{r}_{N^{\bar{\alpha}}}^{\bar{\alpha}} \}$\cite{foulkes2001}.

The backflow transformation replaces every single-particle orbital $\phi_i^{\alpha}(\mathbf{r}_j^{\alpha})$ by a transformed orbital $\phi_i^{\alpha}(\mathbf{x}_j^{\alpha})$, which depends on the position of every electron in the system. Exchanging the coordinates of any two spin-parallel electrons still exchanges two rows of the Slater determinant, so the antisymmetry is preserved. 
The downside is that moving one electron now changes every element of the Slater matrix, preventing the use of efficient rank-1 update formulae and increasing the cost of re-evaluating the determinant by a factor of $N$.
Despite the extra cost, however, the enrichment of the description of correlations between electrons makes SJB wave functions significantly better than SJ wave functions and they are frequently used in VMC and FN-DMC simulations.

The FermiNet \cite{pfau2020} takes the idea of permutation equivariant backflow much further, replacing the orbitals $\phi_i^{\alpha}(\mathbf{r}_j^{\alpha})$ entirely by neural networks.
The orbitals represented by these networks differ from SJB orbitals because they are not functions of a single three-dimensional vector $\mathbf{x}_j^{\alpha}$ but depend in a very general way on $\mathbf{r}_j^{\alpha}$ and all the elements of the sets $\{\mathbf{r}_{/j}^{\alpha}\}$ and $\{\mathbf{r}^{\bar{\alpha}}\}$.
They are best written as $\phi_i^{\alpha}(\mathbf{r}_j^{\alpha}; \{\mathbf{r}_{/j}^{\alpha}\}; \{\mathbf{r}^{\bar{\alpha}}\})$.
The exchange antisymmetry is maintained because $\phi_i^{\alpha}(\mathbf{r}_j^{\alpha}; \{\mathbf{r}_{/j}^{\alpha}\}; \{\mathbf{r}^{\bar{\alpha}}\})$ is totally symmetric on exchange of any pair of coordinates in $\{\mathbf{r}^{\alpha}_{/j}\}$ or $\{\mathbf{r}^{\bar{\alpha}}\}$.
Furthermore, because they are represented as neural networks, the FermiNet orbitals need not be expanded in terms of an explicit basis set, widening the class of functions they can represent \cite{hornik1989}.
In order to build functions with the correct exchange symmetry properties, a carefully constructed neural network architecture is used, which is described below.

The FermiNet architecture consists of two parts: the one-electron stream, which takes electron-nucleus separation vectors $\mathbf{r}^\alpha_i-\mathbf{R}_I$ and distances $|\mathbf{r}^\alpha_i-\mathbf{R}_I|$ as inputs, and the two-electron stream, which takes electron-electron separations $\mathbf{r}^\alpha_i-\mathbf{r}^\beta_j$ and distances $|\mathbf{r}^\alpha_i-\mathbf{r}^\beta_j|$ as inputs, with $i,j\in \{1, 2, \dots, N^\alpha\}$ and $\alpha,\beta \in \{\uparrow, \downarrow\}$.
The inputs to the one-electron stream are concatenated to form one input vector for each electron, and the inputs to the two-electron stream are concatenated to form one input vector for each pair of electrons:
\begin{align}
    \mathbf{h}^{0\alpha}_i &= \left( \mathbf{r}^\alpha_i-\mathbf{R}_I, |\mathbf{r}^\alpha_i-\mathbf{R}_I| \ \forall I \right) ,
    \label{eq:one-electron-stream}
    \\
    \mathbf{h}^{0\alpha\beta}_{ij} &= \left( \mathbf{r}^\alpha_i-\mathbf{r}^\beta_j, |\mathbf{r}^\alpha_i-\mathbf{r}^\beta_j| \right) ,
\end{align}
where the superscript $0$ means that the vectors are the inputs to the first layer of the network.
The distances between particles are passed into the network to help it to model the wave function cusps, \emph{i.e.}, the discontinuities in the derivatives of the wave function when two electrons or an electron and a nucleus coincide.
These discontinuities create divergences in the kinetic energy that exactly cancel the divergences in the potential energy as pairs of charged particles approach each other \cite{pfau2020}.

Each electron stream consists of several layers.
At each layer $l\in\{0, \dots, L-1\}$, the outputs $\mathbf{h}^{l\alpha}_i$ and $\mathbf{h}^{l\alpha\beta}_{ij}$ from the streams are averaged and concatenated in the following way:
\begin{align}
    \mathbf{f}^{l \alpha}_i 
    &= \left( \mathbf{h}^{l\alpha}_i, 
        \mathbf{g}^{l\uparrow}, 
        \mathbf{g}^{l\downarrow}, 
        \mathbf{g}^{l\alpha\uparrow}_i, 
        \mathbf{g}^{l\alpha\downarrow}_i 
    \right), \notag \\
    \mathbf{g}^{l\uparrow} &= \frac{1}{N^\uparrow} \sum^{N^\uparrow}_{j=1} \mathbf{h}^{l\uparrow}_j, 
    \quad 
    \mathbf{g}^{l\downarrow} = \frac{1}{N^\downarrow} \sum^{N^\downarrow}_{j=1} \mathbf{h}^{l\downarrow}_j, \notag \\
    \mathbf{g}^{l\alpha\uparrow}_i &= \frac{1}{N^\uparrow} \sum^{N^\uparrow}_{j=1}\mathbf{h}^{l\alpha\uparrow}_{ij},
    \quad
    \mathbf{g}^{l\alpha\downarrow}_i = \frac{1}{N^\downarrow} \sum^{N^\downarrow}_{j=1} \mathbf{h}^{l\alpha\downarrow}_{ij}.
\end{align}
The concatenated one-electron vectors are then passed into the next layer, as are the two-electron vectors:
\begin{align}
  \mathbf{h}^{(l+1) \alpha}_i &= \tanh(\mathbf{V}^l \mathbf{f}^{l \alpha}_i + \mathbf{b}^l) +\mathbf{h}^{l \alpha}_i, \\
    \mathbf{h}^{(l+1) \alpha\beta}_{ij} &= \tanh(\mathbf{W}^l \mathbf{h}^{l \alpha\beta}_{ij} + \mathbf{c}^l) + \mathbf{h}^{l \alpha\beta}_{ij} ,
\end{align}
where $\mathbf{V}^l$ and $\mathbf{W}^l$ are matrices, $\mathbf{b}^l$ and $\mathbf{c}^l$ are vectors, and all of them are optimizable. We denote the number of hidden units in each layer in the one-electron stream by $n_l$ such that $\mathbf{h}^{l \alpha}_i \in \mathbb{R}^{n_l}, l \in \{0, 1, 2, \dots, L\}$ 
\footnote{
    Although $n_l$ can be different in different layers, in practice, we use an equal number of hidden units in each layer except for the input layer, where $n_0$ is fixed by the number of particles in the system.
}. 
The outputs from the final layer $L$ of the one-electron streams are used to build the many-particle FermiNet orbitals:
\begin{equation}
    \phi^{k\alpha}_i(\mathbf{r}^\alpha_j;\{\mathbf{r}^\alpha_{/j}\};\{\mathbf{r}^{\bar\alpha}\}) 
    = (\mathbf{w}^{k\alpha}_i \cdot \mathbf{h}^{L\alpha}_j + g^{k\alpha}_i) 
    \chi^{k\alpha}_i(\mathbf{r}^\alpha_j),
    \label{eq:ferminet_orbital}
\end{equation}
where $\mathbf{w}^{k\alpha}_i$ is an optimizable vector and $g^{k\alpha}_i$ an optimizable scalar.
The $\chi^{k\alpha}_i(\mathbf{r}^\alpha_j)$ factor is an envelope function to ensure that the wave function satisfies the relevant boundary conditions.
For example, in a system which requires the wave function to tend to zero as $|\mathbf{r}^\alpha_j - \mathbf{R}_m| \rightarrow \infty$, exponential envelopes are used:
\begin{equation}
    \chi^{k\alpha}_i(\mathbf{r}^\alpha_j) =
    \sum_m \pi^{k\alpha}_{im}\exp(-\sigma^{k\alpha}_{im}|\mathbf{r}^\alpha_j - \mathbf{R}_m|),
\end{equation}
where $\pi^{k\alpha}_{im}$ and $\sigma^{k\alpha}_{im}$ are variational parameters. No attempt is made to ensure that the FermiNet orbitals are normalized or orthogonal to each other.

As mentioned earlier, FermiNet orbitals are not functions of one electron position $\mathbf{r}^\alpha_i$ only, but also depend on the positions of all the other electrons in the system in an appropriately permutation invariant way.
No Jastrow factor is needed as the electron-electron correlations are included in the network.
The full wave function is thus a block-diagonal determinant (BD) of the FermiNet orbitals $\phi^{k\alpha}_i(\mathbf{r}^\alpha_j;\{\mathbf{r}^\alpha_{/j}\};\{\mathbf{r}^{\bar\alpha}\})$.
Multiple determinants may also be used, in which case the wave function is a weighted linear combination
\begin{align}
    &\Psi^D_\text{Slater FermiNet}\left(\mathbf{r}^\uparrow_1, \dots, \mathbf{r}^\uparrow_{N^\uparrow}, \mathbf{r}^\downarrow_{1}, \dots,  \mathbf{r}^\downarrow_{N^\downarrow}\right) \notag \\
    &= \sum^D_k \omega_k \det[\phi^{k\uparrow}_{i}(\mathbf{r}^\uparrow_j;\{\mathbf{r}^\uparrow_{/j}\}; \{\mathbf{r}^{\downarrow}\})] \det[\phi^{k\downarrow}_{i}(\mathbf{r}^\downarrow_j;\{\mathbf{r}^\downarrow_{/j}\}; \{\mathbf{r}^{\uparrow}\})],
    \label{eq:block-det-ferminet}
\end{align}
where the superscript $D$ specifies the number of determinants of FermiNet orbitals in the linear combination of determinants that makes up the full wave function, and the ``Slater FermiNet'' subscript serves to specify this specific wave function Ansatz and is discussed in more detail below. In practice, the weights $\omega_k$ are absorbed into the orbitals, which are not normalized.

The VMC method is then applied to the FermiNet Ansatz and the parameters of the network are optimized using a second-order method known as the Kronecker-factored approximate curvature algorithm \cite{pmlr-v37-martens15}.
The aim is to minimize the expectation value of the Hamiltonian $\langle H \rangle$, which acts as our loss function.
For a more detailed explanation of the FermiNet architecture, see Pfau \textit{et al.}\  \cite{pfau2020} and the discussion of the improved JAX implementation \cite{jax2018github} in Spencer \textit{et al.}\ \cite{spencer2020}.

The FermiNet architecture can be extended to study periodic system \cite{gcassella2022,li2022b,wilson2022}.
Consider the basis $\{\mathbf{a_1}, \mathbf{a_2}, \mathbf{a_3}\}$ of the Bravais lattice generated by repeating the finite simulation cell periodically.
Any position vector may be written as $\mathbf{r} = s_1 \mathbf{a_1} + s_2 \mathbf{a_2} + s_3 \mathbf{a_3}$.
To ensure that the FermiNet represents a periodic function, the position coordinates $s_i$ are replaced in the FermiNet inputs by pairs of periodic functions, $s_i \rightarrow (\sin(2\pi s_i), \cos(2\pi s_i))$.
Thus, if any electron is moved by any simulation-cell Bravais lattice vector, the inputs to the network are unchanged. It follows that the output, the value of the wave function, is also unchanged.
A periodic envelope function is used to improve the speed of convergence \cite{gcassella2022}:
\begin{equation}
    \chi^{k\alpha}_i(\mathbf{r}^\alpha_j) 
    = \sum_m \left[ 
        \nu^{k\alpha}_{im}\cos(\mathbf{k}_m \cdot \mathbf{r}^\alpha_j) + \mu^{k\alpha}_{im}\sin(\mathbf{k}_m \cdot \mathbf{r}^\alpha_j)
    \right],
    \label{eq:waves}
\end{equation}
where the $\mathbf{k}_m$ are simulation-cell reciprocal lattice vectors up to the Fermi wavevector of the non-interacting Fermi gas. 
This specific way of adapting the FermiNet to periodic systems was proposed by Cassella \textit{et al.}\ \cite{gcassella2022}, although other similar methods exist \cite{li2022b,wilson2022}.

The FermiNet has only been used to study systems of electrons interacting via Coulomb forces to date, but can easily be adapted to systems of other spin-$1/2$ particles simply by changing the Hamiltonian.
Here we use the periodic FermiNet Ansatz to approximate the ground state of the UFG Hamiltonian in a cubic box subject to periodic boundary conditions.
Since there are no atomic nuclei and the wave function has no electron-nuclear cusps, the inputs to the one-electron streams are simpler than shown in Eq.~(\ref{eq:one-electron-stream}), containing only the particle coordinates
\footnote{
Although, in principle, the particle coordinates $\mathbf{r}^\alpha_i$ are not needed in a periodic system, we have observed that including them often improves convergence and allows the variational Ansatz to achieve a lower energy.
Hence, we have included them in all of our calculations.
}: $\mathbf{h}^{0\alpha}_i = \mathbf{r}^\alpha_i$, with respect to an origin placed at one corner of the simulation cell.
A detailed discussion of translational symmetry of the wave function can be found in section \ref{appendix:translational_invariance} of the Appendix.

As will be demonstrated below, the Slater FermiNet is sufficient to learn the superfluid ground state for small systems but fails for large systems.
Hence, we propose a modification to the method of building orbitals.
The motivation for this modification comes from earlier work using antisymmetrized products of two-particles orbitals known as antisymmetrized geminal power (AGP) wave functions \cite{bouchaud_pair_1988,sorella2004,marchi_resonating_2009,casula_geminal_2003,genovese_general_2020,carlson2003,luo2019}.
We describe the antisymmetrized geminal power singlet (AGPs) wave function in the next section.

The authors of Ref.~\cite{pfau2020} used the term ``FermiNet wave function'' to refer to all wave functions constructed using a FermiNet neural network.
Now that we are going to use almost the same neural network to generate AGPs-based pairing wave functions in addition, more precise terminology is required.
Wave functions of the type introduced in Ref.~\cite{pfau2020}, which contain many-particle generalizations of the one-particle orbitals that appear in Slater determinants, will be called one-determinant or multi-determinant Slater FermiNets.
Wave functions built using determinants of many-particle generalizations of pairing functions will be called one-determinant or multi-determinant AGPs FermiNets. Since every AGPs FermiNet determinant is built using one pairing function or ``geminal'', we also refer to one-geminal or multi-geminal AGPs FermiNets.

\section{Antisymmetrized Geminal Power Wave Function}
\label{section:agp}

The FermiNet and other Ans\"{a}tze that expand the ground state as a linear combination of Slater determinants give very accurate results for many molecules and solids, but may still fail to capture strong two-particle correlations in superfluids.
An alternative starting point, which is better at capturing two-particle correlations, is the antisymmetric geminal power (AGP) wave function \cite{casula_geminal_2003,bajdich_pfaffian_2006,bajdich_generalized_2007,genovese_general_2020}.
This uses an antisymmetrized product of two-particle functions known as pairing orbitals or geminals instead of an antisymmetrized product of single-particle orbitals. 

Although one can build a general AGP wave function with pairings between arbitrary particles, the UFG Hamiltonian only contains interactions between particles of opposite spin.
It is therefore sufficient to consider pairing orbitals involving particles of opposite spin only.
In this case, the wave function is called an antisymmetrized geminal power singlet (AGPs).
The rest of this section summarizes the main features of the AGPs Ansatz and explains how the FermiNet architecture can be modified to produce many-particle generalizations of AGPs pairing orbitals.
Detailed discussions of AGPs wave functions, including derivations of the equations, can be found in Refs.~\cite{casula_geminal_2003,bajdich_pfaffian_2006,bajdich_generalized_2007,genovese_general_2020,luo2019} and the Appendix.

\subsection{AGP Singlet Wave Functions}
It is helpful to start by considering an unpolarized system with an even number ($N=2p$) of particles and total spin $S_z = 0$.
An AGPs wave function for such a system is constructed using a singlet pairing function of the form
\begin{equation}
  \Phi(\mathbf{r}_i,\sigma_i;\mathbf{r}_j,\sigma_j) = \varphi(\mathbf{r}_i,\mathbf{r}_j) \times \langle \sigma_i\sigma_j | \frac{1}{\sqrt{2}} (|\uparrow\downarrow\rangle - |\downarrow\uparrow\rangle ) ,
\end{equation}
where $\varphi(\mathbf{r}_i,\mathbf{r}_j)$ is a symmetric function of its arguments.
We work with spin-assigned wave functions, so we set the spins of particles $1, 2, \dots, p$ to $\uparrow$ and the spins of particles $p+1, p+2, \dots, 2p$ to $\downarrow$.
If, for example, $i\leq p$ and $j>p$, so that particle $i$ is spin-up and particle $j$ is spin-down, the spin-assigned pairing function is
\begin{equation}
  \Phi(\mathbf{r}_i,\uparrow; \mathbf{r}_j,\downarrow) = \varphi(\mathbf{r}_i,\mathbf{r}_j)/\sqrt{2}.
\end{equation}
The spin-assigned singlet pairing function is equal to zero if the spins of particles $i$ and $j$ are the same. 

The spin-assigned AGPs wave function is a determinant of spatial pairing functions \cite{carlson2003,bouchaud_pair_1988}:
\begin{equation}
  \Psi_\text{AGPs}(\mathbf{r}_1^{\uparrow}, \dots, \mathbf{r}_p^{\uparrow}, \mathbf{r}_1^{\downarrow}, \dots, \mathbf{r}_p^{\downarrow})
    = \det[\varphi(\mathbf{r}_i^\uparrow, \mathbf{r}_j^\downarrow)].
    \label{eq:AGPs_det}
\end{equation}
Like all spin-assigned wave functions, it depends on position coordinates only.
For convenience, we have changed the particle labeling scheme: $i$ and $j$ now run from $1$ to $p$ and arrow superscripts have been added to distinguish up-spin from down-spin particles.
Note that the AGPs wave function coincides with the BCS wave function projected onto a fixed particle number subspace (see Ref.~\onlinecite{bouchaud_pair_1988} and Appendix~\ref{appendix:fixed_n_bcs}). It is therefore suitable for describing singlet-paired systems, including $s$-wave superfluids.

\subsection{AGPs with Unpaired States}
We can generalize the spin-assigned AGPs wave function to allow for unpaired particles.
Consider a system with $N = 2p + u + d$ particles, where $p$ is the number of pairs, $u$ is the number of unpaired spin-up particles, and $d$ is the number of unpaired spin-down particles.
The total number of spin-up particles is $p+u$ and the total number of spin-down particles is $p+d$.
The AGPs wave function can be written as a determinant of pairing functions and single-particle orbitals \cite{bajdich_pfaffian_2006,genovese_general_2020,bouchaud_pair_1988} as shown in Eq.~(\ref{eq:agp_uppaired})
\begin{widetext}
\begin{equation}
  \Psi(1,2, \dots,2p+u+d)= \det \left(
    \begin{array}{cccccccc}
      \varphi(\mathbf{r}^\uparrow_1, \mathbf{r}^\downarrow_1)
      & \cdots
      & \varphi(\mathbf{r}^\uparrow_1, \mathbf{r}^\downarrow_{p+d})
      & \phi^{\uparrow}_1(\mathbf{r}^\uparrow_1)
      & \cdots
      & \phi^{\uparrow}_u(\mathbf{r}^\uparrow_1)
      \\
      \varphi(\mathbf{r}^\uparrow_2, \mathbf{r}^\downarrow_1)
      & \cdots
      & \varphi(\mathbf{r}^\uparrow_2, \mathbf{r}^\downarrow_{p+d})
      & \phi^{\uparrow}_1(\mathbf{r}^\uparrow_2)
      & \cdots
      & \phi^{\uparrow}_u(\mathbf{r}^\uparrow_2)
      \\
      \vdots
      & \ddots
      & \vdots
      & \vdots
      & \ddots
      & \vdots 
      \\
      \varphi(\mathbf{r}^\uparrow_{p+u}, \mathbf{r}^\downarrow_1)
      & \cdots
      & \varphi(\mathbf{r}^\uparrow_{p+u}, \mathbf{r}^\downarrow_{p+d})
      & \phi^{\uparrow}_1(\mathbf{r}^\uparrow_{p+u})
      & \cdots
      & \phi^{\uparrow}_u(\mathbf{r}^\uparrow_{p+u})
      \\
      \phi^{\downarrow}_1(\mathbf{r}^\downarrow_1)
      & \cdots
      & \phi^{\downarrow}_1(\mathbf{r}^\downarrow_{p+d})
      & 0
      & \cdots
      & 0 
      \\
      \phi^{\downarrow}_2(\mathbf{r}^\downarrow_1)
      & \cdots
      & \phi^{\downarrow}_2(\mathbf{r}^\downarrow_{p+d})
      & 0
      & \cdots
      & 0 
      \\
      \vdots
      & \ddots
      & \vdots
      & \vdots
      & \ddots
      & \vdots 
      \\
      \phi^{\downarrow}_d(\mathbf{r}^\downarrow_1)
      & \cdots
      & \phi^{\downarrow}_d(\mathbf{r}^\downarrow_{p+d})
      & 0
      & \cdots
      & 0 
    \end{array} 
  \right) ,
  \label{eq:agp_uppaired}
\end{equation}
\end{widetext}
where $\varphi(\mathbf{r}^\uparrow_i, \mathbf{r}^\downarrow_j)$ is an arbitrary singlet pairing function and $\phi^{\sigma_i}_i(\mathbf{r}^{\sigma_i}_j)$ are arbitrary single-particle functions.
For the UFG considered in this paper, we only need the case where $u=1$ and $d=0$ or vice versa.
This represents a fully paired $2p$-particle system to which one particle has been added.

\section{AGP Singlet FermiNet}
\label{section:agps_ferminet}

Having discussed the form of the AGPs wave function, we now discuss how it can be implemented using FermiNet.
In the original Slater FermiNet architecture, the outputs of the one-electron stream are used to build FermiNet orbitals $\phi^{k\alpha}_{i}(\mathbf{r}^\alpha_j;\{\mathbf{r}^\alpha_{/j}\}; \{\mathbf{r}^{\bar\alpha}\})$.
The full many-particle wave function is a weighted sum of terms, each of which is the product of one up-spin and one down-spin determinant of the FermiNet orbital matrices, as shown in Eq.~(\ref{eq:block-det-ferminet}).

To build a many-particle pairing function using the neural-network part of FermiNet, one can make use of its outputs $\mathbf{h}^{L\alpha}_i$ from the last layer $L$ of the one-electron stream.
Instead of using these outputs to build FermiNet orbitals, as in Eq.~(\ref{eq:ferminet_orbital}), they can be used to build FermiNet pairing orbitals, also known as FermiNet geminals:
\begin{multline}
    \varphi^{k}(\mathbf{r}^\alpha_i, \mathbf{r}^{\bar\alpha}_j;
      \{\mathbf{r}^\alpha_{/i}\};\{\mathbf{r}^{\bar\alpha}_{/j}\}) \\
    = [\mathbf{w}^{k} \cdot (\mathbf{h}^{L\alpha}_i \odot 
      \mathbf{h}^{L\bar\alpha}_j) + g^k] \chi^{k}(\mathbf{r}^\alpha_i)
      \chi^{k}(\mathbf{r}^{\bar\alpha}_j),
    \label{eq:ggno2e}
\end{multline}
where $\chi^k(\mathbf{r})$ are the envelope functions, $\mathbf{w}^{k}$ are vectors, $g^k$ a scalar, and $\odot$ denotes the element-wise product.
Note that the same FermiNet geminal is used for all pairs of particles, so the envelope functions in Eq.~(\ref{eq:ggno2e}) do not require the particle and spin indices that appear in the envelope functions of the FermiNet orbitals defined in Eq.~(\ref{eq:ferminet_orbital}).
This construction generates a many-particle pairing function between particles $\mathbf{r}^\alpha_i$ and $\mathbf{r}^{\bar\alpha}_j$, retaining the permutation invariant property possessed by FermiNet orbitals.
Depending on the number of FermiNet geminals generated, the wave function can be written as one or a weighted sum of multiple determinants of FermiNet geminals,
\begin{align}
    &\Psi^{D}_\text{AGPs FermiNet}\left(\mathbf{r}^\uparrow_1, \dots, \mathbf{r}^\downarrow_{N^\downarrow} \right) \notag \\
    &= \sum^D_k \omega_k \det[\varphi^{k}(\mathbf{r}^\alpha_i, \mathbf{r}^{\bar\alpha}_j; \{\mathbf{r}^\alpha_{/i}\};\{\mathbf{r}^{\bar\alpha}_{/j}\})] ,
    \label{eq:agps_ferminet}
\end{align}
where the superscript $D$ in $\Psi^{D}_\text{AGPs FermiNet}$ specifies the number of determinants (and thus FermiNet geminals) appearing in the linear combination that makes up the wave function.
This is analogous to a weighted sum of conventional single-determinant AGPs wave functions of the type defined in Eq.~(\ref{eq:AGPs_det}), but the replacement of the two-particle pairing orbitals by FermiNet geminals that depend on the positions of all the particles makes it much more general.

Although using the outputs from the one-electron stream is sufficient to build an AGPs, one can also include the outputs from the two-electron stream:
\begin{align}
    \varphi^{k}&(\mathbf{r}^\alpha_i, \mathbf{r}^{\bar\alpha}_j;
      \{\mathbf{r}^\alpha_{/i}\};\{\mathbf{r}^{\bar\alpha}_{/j}\}) & \notag \\
    = [&\mathbf{w}^{k}_1 \cdot (\mathbf{h}^{L\alpha}_i \odot 
      \mathbf{h}^{L\bar\alpha}_j) \notag \\
    &+ \mathbf{w}^{k}_2 \cdot 
      (\mathbf{h}^{L\alpha\bar\alpha}_{ij} \odot 
      \mathbf{h}^{L\bar\alpha\alpha}_{ji}) + g^k]
      \chi^{k}(\mathbf{r}^\alpha_i)\chi^{k}(\mathbf{r}^{\bar\alpha}_j) ,
    \label{eq:ggw2e}
\end{align}
where $\mathbf{w}^{k}_1$ and $\mathbf{w}^{k}_2$ are vectors.

Note that Eqs.~(\ref{eq:ggno2e}) and (\ref{eq:ggw2e}) are two possible ways of building a many-particle pairing function.
There are many others ways and they are all valid as long as the appropriate symmetries are preserved.
An alternative method is given by Xie \emph{et al.} \cite{xie2022}.

The benefit of building AGPs-like wave functions using the FermiNet is that the many-particle pairing function $\varphi(\mathbf{r}^\alpha_i, \mathbf{r}^{\bar\alpha}_j;\{\mathbf{r}^\alpha_{/i}\};\{\mathbf{r}^{\bar\alpha}_{/j}\})$ now depends not only on $\mathbf{r}_i^{\alpha}$ and $\mathbf{r}_j^{\bar{\alpha}}$ but also on the positions of the other particles in the system
\footnote{
  Note that the idea of an AGPs/BCS FermiNet is very similar to the previous work by Luo and Clark \cite{luo2019}, where they used the neural network backflow (NNB) wave function implemented on top of a Bogoliubov-de Gennes/BCS wave function to study lattice systems.
}.
Correlations between the singlet pair and the other particles can thus be captured.
In a similar way, the original Slater FermiNet replaced Hartree-Fock-like single-particle orbitals $\phi_i^{\alpha}(\mathbf{r}_j^{\alpha})$ by many-particle orbitals (FermiNet orbitals) $ \phi^{\alpha}_{i}(\mathbf{r}^\alpha_j;\{\mathbf{r}^\alpha_{/j}\}; \{\mathbf{r}^{\bar\alpha}\})$, helping to capture correlations between the particle at $\mathbf{r}_j^{\alpha}$ and all other particles.

\subsection{Relations between the Slater FermiNet and the AGPs FermiNet}
Next, we clarify the relation between the Slater FermiNet with block-diagonal determinants and the AGPs FermiNet, showing that the AGPs FermiNet is the more general of the two.
A FermiNet geminal with a two-particle stream term is even more general than a FermiNet geminal without, so it is sufficient for this purpose to omit the two-particle stream term.
We also neglect the envelope functions and the bias term, $g^k$, which is set to 0 in all results presented here.
The use of envelope functions circumvents numerical difficulties in finite systems and can speed up the network optimization, but does not affect the generality of the Ansatz.

Let us first define a many-particle pairing function in the following way:
\begin{equation}
    \varphi(\mathbf{r}^\uparrow_i, \mathbf{r}^\downarrow_j; 
      \{\mathbf{r}^\uparrow_{/i}\};\{\mathbf{r}^{\downarrow}_{/j}\})
    = \sum_{kl} W_{kl} h^{L\uparrow (k)}_i h^{L\downarrow(l)}_j ,
    \label{eq:ggij}
\end{equation}
where $h^{L\alpha (k)}_i = [\mathbf{h}^{L\alpha}_i]_k$ are the outputs from the final layer of the one-electron stream for particle $i$ of spin $\alpha$.
As we explain below, Eq.~(\ref{eq:ggij}) is equivalent to the simpler FermiNet geminal described above:
\begin{equation}
    \varphi(\mathbf{r}^\uparrow_i, \mathbf{r}^{\downarrow}_j; 
      \{\mathbf{r}^\uparrow_{/i}\};\{\mathbf{r}^{\downarrow}_{/j}\})
    = \mathbf{w} \cdot (\mathbf{h}^{L\uparrow}_i \odot \mathbf{h}^{L\downarrow}_j).
    \label{eq:ggwob}
\end{equation}
We choose to write the many-particle pairing function in the form of Eq.~(\ref{eq:ggij}) only because this makes it easier to relate to FermiNet orbitals.
Since Eqs.~(\ref{eq:ggij}) and (\ref{eq:ggwob}) are equivalent, the choice does not affect the conclusions of the argument.
In the rest of this section, for the sake of simplicity, we omit the sets $\{\mathbf{r}_{/i}^{\uparrow}\}$ and $\{\mathbf{r}_{/j}^{\downarrow}\}$ from the arguments of the many-particle pairing functions and orbitals.

To explain the equivalence of Eqs.~(\ref{eq:ggij}) and (\ref{eq:ggwob}), it is helpful to represent the matrix $W_{kl}$ as its singular-value decomposition (SVD):
\begin{equation}
    W_{kl} = \sum^{n_L}_{\gamma=1} \sigma_\gamma U_{\gamma k} V_{\gamma l},
    \label{eq:svd}
\end{equation}
where $U \in \mathbb{R}^{n_L \times n_L}$ and $V  \in \mathbb{R}^{n_L \times n_L}$ are orthogonal matrices and $n_L$ is the size of the vectors $\mathbf{h}_i^{L\alpha}$ output by the final layer $L$ of the one-electron stream.
This is also known as the number of hidden units in layer $L$.
The many-particle pairing function in Eq.~(\ref{eq:ggij}) becomes
\begin{align}
    \varphi(\mathbf{r}^\uparrow_i, \mathbf{r}^\downarrow_j)
    &= \sum_{kl} W_{kl} h^{L\uparrow (k)}_i h^{L\downarrow(l)}_j \\
    &= \sum_\gamma \sigma_\gamma \left(\sum_k U_{\gamma k} h^{L\uparrow (k)}_i\right) 
        \left(\sum_l V_{\gamma l} h^{L\uparrow (l)}_j\right) \\
    &= \sum_\gamma \sigma_\gamma (U \mathbf{h}^{L\uparrow}_i)_\gamma 
        (V \mathbf{h}^{L\downarrow}_j)_\gamma.
\end{align}
Given the universal approximation theorem \cite{hornik1989}, and the fact that every layer of the network contains an arbitrary linear transformation, it is reasonable to assume that the functions $\mathbf{h}^{L\alpha}_i$ and $O\mathbf{h}^{L\alpha}_i$, where $O$ is $U$ or $V$, have the same variational freedom and information content.
In other words, we assume that any network capable of representing $\mathbf{h}^{L\alpha}_i$ can also represent $O\mathbf{h}^{L\alpha}_i$, since this is merely a rotation of the vectors $\mathbf{h}^{L\alpha}_i$ in the last layer.
We thus define $\tilde{\mathbf{h}}^{L\uparrow}_i = U \mathbf{h}^{L\uparrow}_i$ and $\tilde{\mathbf{h}}^{L\downarrow}_i = V \mathbf{h}^{L\downarrow}_i$, such that the many-particle pairing function becomes
\begin{align}
    \varphi(\mathbf{r}^\uparrow_i, \mathbf{r}^\downarrow_j)
    &= \sum_\gamma \sigma_\gamma (U \mathbf{h}^{L\uparrow}_i)_\gamma 
        (V \mathbf{h}^{L\downarrow}_j)_\gamma \notag \\
    &= \sum_\gamma \sigma_\gamma (\tilde{\mathbf{h}}^{L\uparrow}_i 
        \odot \tilde{\mathbf{h}}^{L\downarrow}_j)_{\gamma},
\end{align}
which is equivalent to Eq.~(\ref{eq:ggwob}).

To relate the AGPs FermiNet and the Slater FermiNet, we expand an AGPs determinant constructed using the many-particle pairing function from Eq.~(\ref{eq:ggij}) as a sum of block-diagonal determinants of FermiNet orbitals.
It will be sufficient to consider matrices $W \in \mathbb{R}^{n_L \times n_L}$ of rank $M$, with $p\le M\le n_L$.
We can decompose any such matrix using rank factorization,
\begin{equation}
    W_{kl} = \sum^M_{\gamma=1} F_{\gamma k} G_{\gamma l},
    \label{eq:rf}
\end{equation}
where $F$ and $G$ are matrices in $\mathbb{R}^{M\times n_L}$.
Equation~(\ref{eq:ggij}) then becomes
\begin{align}
    \varphi(\mathbf{r}^\uparrow_i, \mathbf{r}^\downarrow_j)
    &= \sum_{\gamma=1}^{M} \left(\sum_{k=1}^{n_L} F_{\gamma k} h^{L\uparrow (k)}_i\right) \left(\sum_{l=1}^{n_L} G_{\gamma l}  h^{L\downarrow (l)}_j\right) \notag \\
    &= \sum_{\gamma=1}^{M} \phi^\uparrow_\gamma(\mathbf{r}^\uparrow_i) \phi^\downarrow_\gamma(\mathbf{r}^\downarrow_j),
\end{align}
where the last line defines the functions $\phi_{\gamma}^{\uparrow}$ and $\phi_{\gamma}^{\downarrow}$.
In the case when $M = p$, where $p = N/2$ is the number of pairs in the system, the determinant of the many-particle pairing function can be written as a block-diagonal determinant of FermiNet orbitals:
\begin{align}
    \det[\varphi(\mathbf{r}^\uparrow_i, \mathbf{r}^\downarrow_j)]
    &= \det[\sum_{\gamma=1}^p \phi^\uparrow_\gamma(\mathbf{r}^\uparrow_i) \phi^\downarrow_\gamma(\mathbf{r}^\downarrow_j)] \notag \\
    &= \det[\left(\phi^{\uparrow T} \phi^{\downarrow}\right)_{ij}] \notag \\
    &= \det[\phi^\uparrow_i(\mathbf{r}^\uparrow_j)] \det[\phi^\downarrow_i(\mathbf{r}^\downarrow_j)]
    \label{eq:M_equals_p}
\end{align}
where $[\phi^{\alpha}]_{ij} = \phi^\alpha_i(\mathbf{r}^\alpha_j)$ are matrices in $\mathbb{R}^{M\times p}$ with $M=p$.
The product of two $p\times p$ determinants can be written as the determinant of a single $2p \times 2p$ matrix, with the $p\times p$ spin-up and spin-down blocks on the diagonal.
Therefore, a single-geminal AGPs FermiNet wave function constructed using the many-particle pairing function from Eq.~(\ref{eq:ggij}) with a rank-$p$ matrix $W_{kl}$ is equivalent to a $2p \times 2p$ block-diagonal determinant of FermiNet orbitals.
The equivalence is already well known \cite{marchi_resonating_2009} for AGPs wave functions constructed using conventional two-particle orbitals.

Now consider the more general case where $p\le M\le n_L$.
The Cauchy-Binet formula states that
\begin{widetext}
\begin{align}
  &\det[\varphi(\mathbf{r}^\uparrow_i, \mathbf{r}^\downarrow_j)]
  = \det[\sum_{\gamma=1}^{M} \phi^\uparrow_\gamma(\mathbf{r}^\uparrow_i) \phi^\downarrow_\gamma(\mathbf{r}^\downarrow_j)] \notag \\
  &= \sum_{1 \le j_1 < j_2 < \cdots < j_p \le M}
      \det
      \begin{pmatrix}
          \phi^\uparrow_{j_1}(\mathbf{r}^\uparrow_1) 
          & \phi^\uparrow_{j_1}(\mathbf{r}^\uparrow_2) 
          & \cdots & \phi^\uparrow_{j_1}(\mathbf{r}^\uparrow_p)  
          \\
          \phi^\uparrow_{j_2}(\mathbf{r}^\uparrow_1) 
          & \phi^\uparrow_{j_2}(\mathbf{r}^\uparrow_2) 
          & \cdots & \phi^\uparrow_{j_2}(\mathbf{r}^\uparrow_p)  
          \\
          \vdots & \vdots & \ddots & \vdots
          \\
          \phi^\uparrow_{j_p}(\mathbf{r}^\uparrow_1) 
          & \phi^\uparrow_{j_p}(\mathbf{r}^\uparrow_2) 
          & \cdots & \phi^\uparrow_{j_p}(\mathbf{r}^\uparrow_p) 
      \end{pmatrix}
      \det
      \begin{pmatrix}
          \phi^\downarrow_{j_1}(\mathbf{r}^\downarrow_1) 
          & \phi^\downarrow_{j_1}(\mathbf{r}^\downarrow_2) 
          & \cdots & \phi^\downarrow_{j_1}(\mathbf{r}^\downarrow_p)  
          \\
          \phi^\downarrow_{j_2}(\mathbf{r}^\downarrow_1) 
          & \phi^\downarrow_{j_2}(\mathbf{r}^\downarrow_2) 
          & \cdots & \phi^\downarrow_{j_2}(\mathbf{r}^\downarrow_p)  
          \\
          \vdots & \vdots & \ddots & \vdots
          \\
          \phi^\downarrow_{j_p}(\mathbf{r}^\downarrow_1) 
          & \phi^\downarrow_{j_p}(\mathbf{r}^\downarrow_2) 
          & \cdots & \phi^\downarrow_{j_p}(\mathbf{r}^\downarrow_p)  
      \end{pmatrix},
\end{align}
\end{widetext}
where the sum is over all $\binom{M}{p}$ distinct choices of $p$ rows from the two $M \times p$ matrices $\phi^\uparrow_\gamma(\mathbf{r}^\uparrow_i)$ and $\phi^\downarrow_\gamma(\mathbf{r}^\downarrow_i)$.
The products of the determinants of the two $p \times p$ matrices associated with each such choice are summed to reproduce the AGPs.
This is similar to the linear combination of multiple block-diagonal-determinants of FermiNet orbitals without weights given by Eq.~(\ref{eq:block-det-ferminet}) and in the original FermiNet paper \cite{pfau2020}
\footnote{
  To get a rough estimate of the number of terms in the sum, we can take the number of hidden units in the one-electron stream to be $M = n_L \sim 200$ and the number of pairs to be $p \sim 20$, this number is approximately $10^{27}$.
}.

Note that the intermediate layers, \emph{i.e.}, the one and two-electron streams, are identical in the Slater FermiNet and the AGPs FermiNet.
The only modifications are made at the orbital shaping layer, or, equivalently, the method of antisymmetrization has changed.
Thus, the representational power of the intermediate layers of the AGPs FermiNet remains the same as for the Slater FermiNet. Thus, it must be the method of antisymmetrization that limits the performance of the Slater FermiNet when applied to the UFG.

We have shown that a single AGPs determinant constructed using the many-particle pairing function from Eq.~(\ref{eq:ggij}) with a matrix $W_{kl}$ of rank greater than $p$ contains multiple block-diagonal determinants of FermiNet orbitals.
If the rank of $W_{kl}$ is equal to $p$, the AGPs is equivalent to a single block-diagonal FermiNet determinant.
Conversely, any single-determinant FermiNet wave function can be written as an AGPs of rank $p$.
Therefore, the AGPs FermiNet provides a more powerful Ansatz with fewer variational parameters than the Slater FermiNet, since the former contains the latter.

It is worth mentioning another advantage of using the FermiNet to build geminals.
By generating more sets of independent parameters $\mathbf{w}^{k}_i$ in Eq.~(\ref{eq:ggw2e}), one can easily construct an arbitrary number $N_{\det}$ of FermiNet geminals $\varphi^{k}(\mathbf{r}^\alpha_i, \mathbf{r}^{\bar\alpha}_j;
\{\mathbf{r}^\alpha_{/i}\};\{\mathbf{r}^{\bar\alpha}_{/j}\})$ with $k \in \{1, 2, \dots, N_{\det}\}$, all without the use of a basis set.
This allows one to use weighted sums of AGPs determinants as trial wave functions, similar to the weighted sum of conventional FermiNet determinants seen in Eq.~(\ref{eq:block-det-ferminet}).

\subsection{AGPs FermiNet with Unpaired States}
To extend the AGPs FermiNet to systems with unpaired states, such as an odd-number of particle system, we use FermiNet geminals and orbitals to replace both the pairing orbitals and the single-particle orbitals in Eq.~(\ref{eq:agp_uppaired}).
In this work we consider systems with equal numbers of up-spin and down-spin particles, which are assumed to be fully paired, and systems containing one additional unpaired particle, which may have spin up or spin down. 
For example, the AGPs FermiNet with an extra spin-up particle is given by
\begin{widetext}
\begin{align}
    &\Psi^\text{AGPs}_\text{FermiNet}\left(\mathbf{r}^\uparrow_1, \dots, \mathbf{r}^\uparrow_{p+1}, \mathbf{r}^\downarrow_{1}, \dots,  \mathbf{r}^\downarrow_{p}\right) \notag \\
    &= \sum_k \omega_k \det \left(
    \begin{array}{cccc}
        \varphi^k(\mathbf{r}^\uparrow_1, \mathbf{r}^\downarrow_1;
        \{\mathbf{r}^\uparrow_{/1}\};\{\mathbf{r}^{\downarrow}_{/1}\})
        & \cdots
        & \varphi^k(\mathbf{r}^\uparrow_1, \mathbf{r}^\downarrow_{p};
        \{\mathbf{r}^\uparrow_{/1}\};\{\mathbf{r}^{\downarrow}_{/p}\})
        & \phi^{k\uparrow}_1(\mathbf{r}^\uparrow_1;\{\mathbf{r}^\uparrow_{/1}\}; \{\mathbf{r}^{\downarrow}\})
        \\
        \varphi^k(\mathbf{r}^\uparrow_2, \mathbf{r}^\downarrow_1;
        \{\mathbf{r}^\uparrow_{/2}\};\{\mathbf{r}^{\downarrow}_{/1}\})
        & \cdots
        & \varphi^k(\mathbf{r}^\uparrow_2, \mathbf{r}^\downarrow_{p};
        \{\mathbf{r}^\uparrow_{/2}\};\{\mathbf{r}^{\downarrow}_{/p}\})
        & \phi^{k\uparrow}_1(\mathbf{r}^\uparrow_2;\{\mathbf{r}^\uparrow_{/2}\}; \{\mathbf{r}^{\downarrow}\})
        \\
        \vdots
        & \ddots
        & \vdots
        & \vdots
        \\
        \varphi^k(\mathbf{r}^\uparrow_{p+1}, \mathbf{r}^\downarrow_1;
        \{\mathbf{r}^\uparrow_{/p+1}\};\{\mathbf{r}^{\downarrow}_{/1}\})
        & \cdots
        & \varphi^k(\mathbf{r}^\uparrow_{p+1}, \mathbf{r}^\downarrow_{p};
        \{\mathbf{r}^\uparrow_{/p+1}\};\{\mathbf{r}^{\downarrow}_{/p}\})
        & \phi^{k\uparrow}_1(\mathbf{r}^\uparrow_{p+1};\{\mathbf{r}^\uparrow_{/p+1}\}; \{\mathbf{r}^{\downarrow}\})
    \end{array}
  \right) .
  \label{eq:agpsferminet_uppaired}
\end{align}
\end{widetext}
where $\varphi^k(\mathbf{r}^\uparrow_i, \mathbf{r}^\downarrow_j; \{\mathbf{r}^\uparrow_{/i}\};\{\mathbf{r}^{\downarrow}_{/j}\})$ can either be defined as Eq.~(\ref{eq:ggno2e}) or Eq.~(\ref{eq:ggw2e}), and $\phi^{k\uparrow}_i(\mathbf{r}^\uparrow_{j};\{\mathbf{r}^\uparrow_{/j}\}; \{\mathbf{r}^{\downarrow}\})$ is defined in Eq.~(\ref{eq:ferminet_orbital}).
In practice, we generate the required number of FermiNet geminal and orbital parameters in batch at the orbital projection layer.
For example, one FermiNet geminal and one FermiNet orbital are generated per determinant for a system with one extra spin-up particle.

\section{Results}
\label{section:results}

The power of the AGPs FermiNet Ansatz may be demonstrated by studying the UFG.
The Hamiltonian is 
\begin{equation}
    \hat{H} = -\frac{1}{2} \sum^N_i \nabla^2_i + \sum^{N/2}_{ij} U(\mathbf{r}^\uparrow_i - \mathbf{r}^{\downarrow}_j),
\end{equation}
where 
\begin{equation}
    U(\mathbf{r}) = - \frac{2v_0 \mu^2}{\cosh^2(\mu r)}
\end{equation}
is the modified P\"oschl-Teller potential, which is widely used in variational and diffusion QMC simulations
\cite{carlson2003,chang_quantum_2004,morris_ultracold_2010,chang2005,gezerlis_low-density_2010,forbes2011,he_superfluid_2020,song2021}.
It would be preferable to use a delta function interaction with an infinite $s$-wave scattering length, but it is difficult to simulate systems with delta-like potentials using QMC methods.
Thus, a finite but short-ranged interaction is typically used.
The $s$-wave scattering length of the P\"oschl-Teller potential diverges when $v_0 = 1$.
By changing the value of $\mu$ at fixed $v_0 = 1$, it is possible to vary the effective range of the interaction, $r_e = 2/\mu$, whilst holding the $s$-wave scattering length infinite.

We choose to study a system with density parameter $r_s=1$, where $r_s$, the radius of a sphere that contains one particle on average, provides a convenient measure of the inter-particle distance.
Throughout this work, we employ the dimensionless system based on Hartree atomic units: the unit of length is the Bohr radius, $a_0$, and the unit of energy is the Hartree.
To ensure that the range of the interaction is small compared with the inter-particle separation, we set $\mu = 12$ ($r_e = 1/6$), keeping $v_0 = 1$ to ensure that the scattering length remains infinite
\footnote{
  Although the range of the potential is finite, which introduces a second length scale even at unitarity, most of the results in the literature have been obtained with the same value of $\mu$.
  This allows us to compare our results with theirs.
  To obtain exact results, one has to study the limit as $\mu \rightarrow \infty$, but converging the simulations becomes more difficult as $\mu$ increases.
}.
We have also simulated the system with $k_F = 1$ (equivalent to $r_s = (9\pi/4)^{1/3} \approx 1.92$) and $\mu=12$ to compare with the fixed-node diffusion Monte Carlo (FN-DMC) result from Forbes \textit{et al.}~\cite{forbes2011}.

We use both the Slater FermiNet with multiple block-diagonal determinants and the AGPs FermiNet with multiple geminals to study the unitary Fermi gas from $N=4$ to $N=38$ particles in a cubic box subject to periodic boundary conditions, as well as AGPs FermiNet on the $N=66$ system
\footnote{
  As previous works \cite{he_superfluid_2020,carlson2011auxiliary,forbes2012effective,morris_ultracold_2010} have shown, the ground-state energy of the $N=66$ particle UFG is the closest to the ground-state energy in the thermodynamic limit even among larger systems.
  This can also be seen by comparing the ratio of $E_{FG}$ and the energy per particle from summing over the energies occupied by the particles in the discretized $k$-space, where this ratio is just $0.5\%$ away from unity for $N=66$. 
}.
The same network size, number of determinants, and number of training iterations are used for both Ans\"atze.
The FermiNet orbitals are given by Eq.~(\ref{eq:ferminet_orbital}) without the bias term.
The FermiNet geminal used for systems containing from $4$ to $28$ particles is the one defined in Eq.~(\ref{eq:ggno2e}).
Unless otherwise stated, all calculations used a linear combination of 32 determinants or 32 geminals. 
Including contributions from the two-electron stream improves the optimization rate and can achieve a slightly lower variational energy in larger systems, so  Eq.~(\ref{eq:ggw2e}) was used for systems of $N\ge29$. 
The inclusion of plane-wave envelopes as defined in Eq.~(\ref{eq:waves}) also improves the optimization rate.
For molecular and electron gas systems, we have found that the bias term in the FermiNet orbital projection (Eq.~\ref{eq:ferminet_orbital}) does not affect the accuracy or optimization of the model. We hence set the bias term, $g^{k\alpha}_i$ and $g^k$ as appropriate, to zero for all calculations presented here.

\begin{figure*}[ht!]
    \centering
    \begin{subfigure}[b]{0.45\textwidth}
        \centering
        \includegraphics[width=\textwidth]{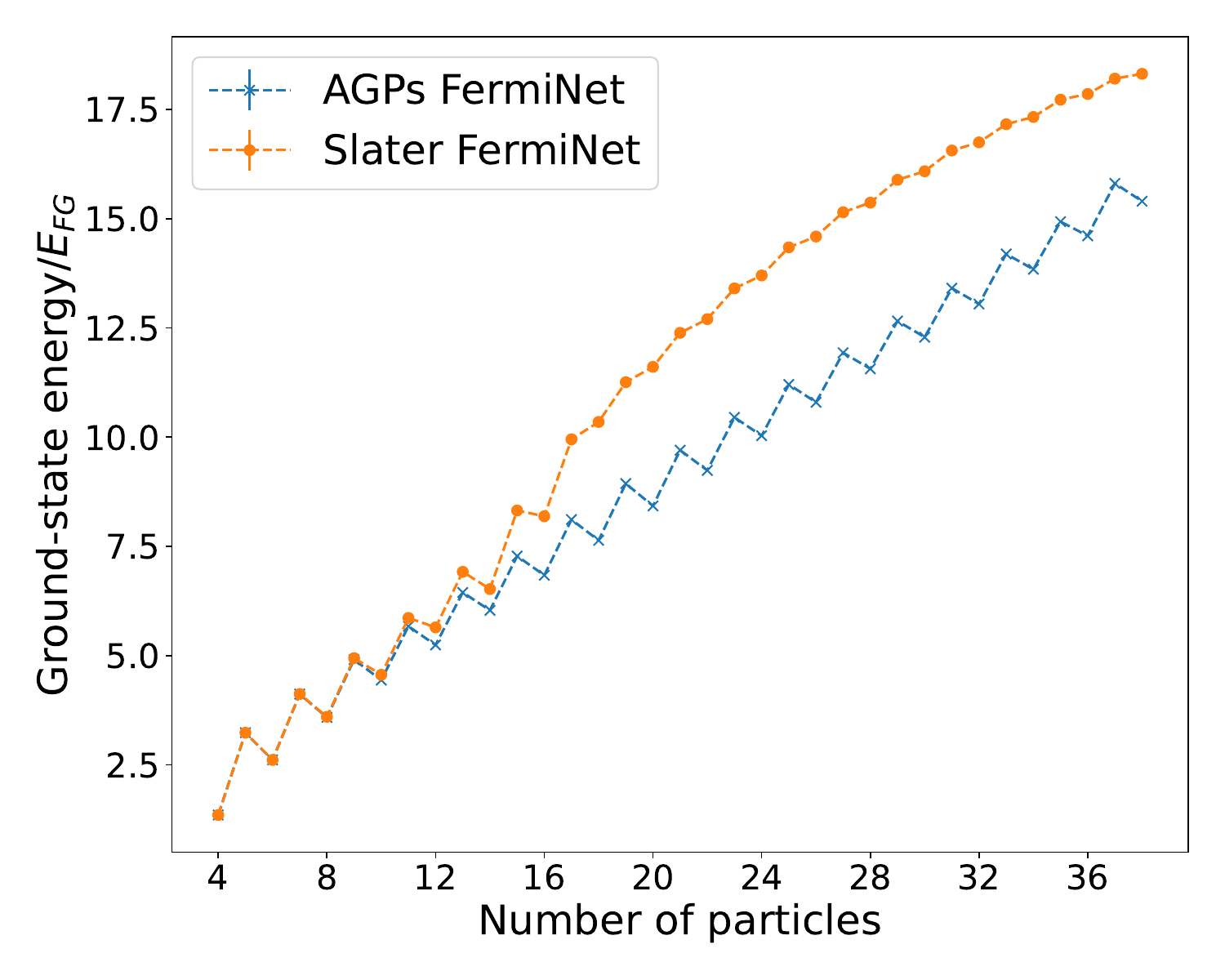}
        \caption{The total energy of the UFG simulation cell, measured in units of the free Fermi gas energy $E_\text{FG}$. The Slater FermiNet Ansatz begins to fail when $N \gtrapprox 10$.}
        \label{fig:e_n}
    \end{subfigure}
    \begin{subfigure}[b]{0.45\textwidth}
        \centering
        \includegraphics[width=\textwidth]{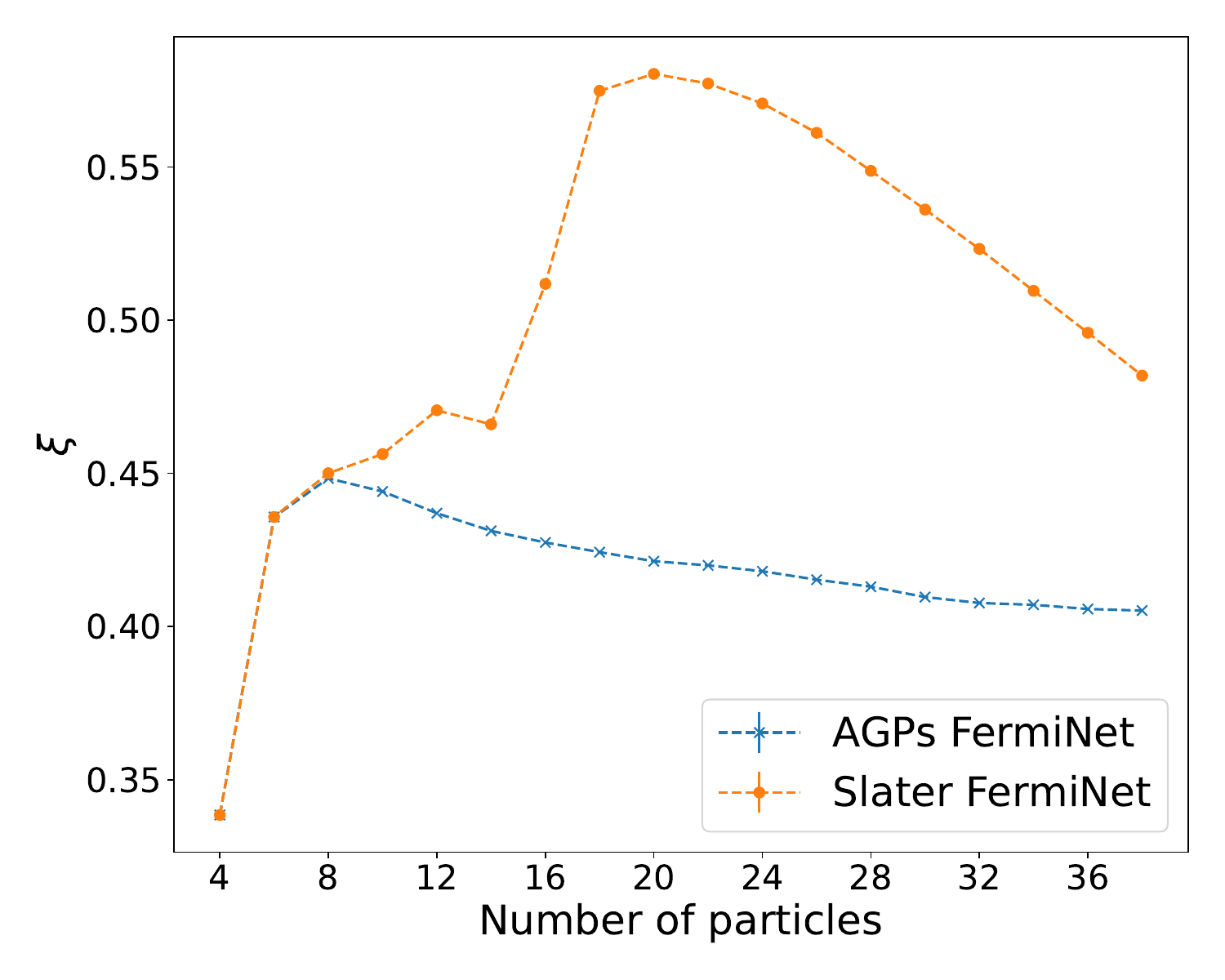}
        \caption{The Bertsch parameter $\xi$ (the ratio of the interacting and non-interacting ground-state energies per particle) as a function of the number of particles $N$.}
        \label{fig:xi}
    \end{subfigure}
    \caption{Comparison between results obtained using the AGPs FermiNet and the Slater FermiNet for different numbers of particles, $N$, with $r_s=1$ and $\mu = 12$.
    All simulations used 32 determinants, 300,000 optimization steps, and the same hyperparameters, which are detailed in the Appendix.}
\end{figure*}

A comparison of the ground-state energy expectation values given by the two Ans\"atze is shown in Fig.~\ref{fig:e_n}.
The Slater FermiNet, which consists of a linear combination of block-diagonal determinants of FermiNet orbitals, performs well when the number of particles $N$ is smaller than around $10$, but the AGPs FermiNet is much superior in larger systems.
It is clear that the Slater FermiNet Ansatz has difficulties learning the ground states of large paired systems
\footnote{
  All calculations presented in this section are performed using the Slater FermiNet with \emph{block-diagonal determinants}. 
  A small set of calculations are performed with the Slater FermiNet with \emph{dense determinants}, which are presented in the Appendix section (Appendix \ref{appendix:dense}).
  In short, we did found slight improvements over the block-diagonal Slater FermiNet results, but it is still qualitatively wrong and is also much worse than the AGPs FermiNet results.
}.

In systems containing an odd number of particles, one must be left unpaired.
This raises the energy a little and explains the zigzag shape of Fig.~\ref{fig:e_n}.
The odd-even staggering is lost for larger systems with the Slater FermiNet Ansatz, indicating the absence of pair formation \cite{carlson2003,palkanoglou2020}.
The Slater FermiNet fails to learn the superfluid state.
For the AGPs FermiNet, by contrast, the amplitude of the odd-even zigzag remains constant, superposed on the linear increase with $N$ expected of any extensive quantity.

Another comparison between the two Ans\"atze is shown in Fig.~\ref{fig:xi}, which depicts the ratio of the interacting and non-interacting energies per particle, known as the Bertsch parameter \cite{bertsch1999} and defined in Eq.~(\ref{eq:bertsch}), as a function of $N$.
All FermiNet energies are variational and the non-interacting energies are exact, so the AGPs FermiNet, for which the Bertsch parameter is lower by up to around 30\%, is much the better of the two Ans\"{a}tze. 

\begin{figure}[ht!]
    \centering
    \includegraphics[width=0.9\columnwidth]{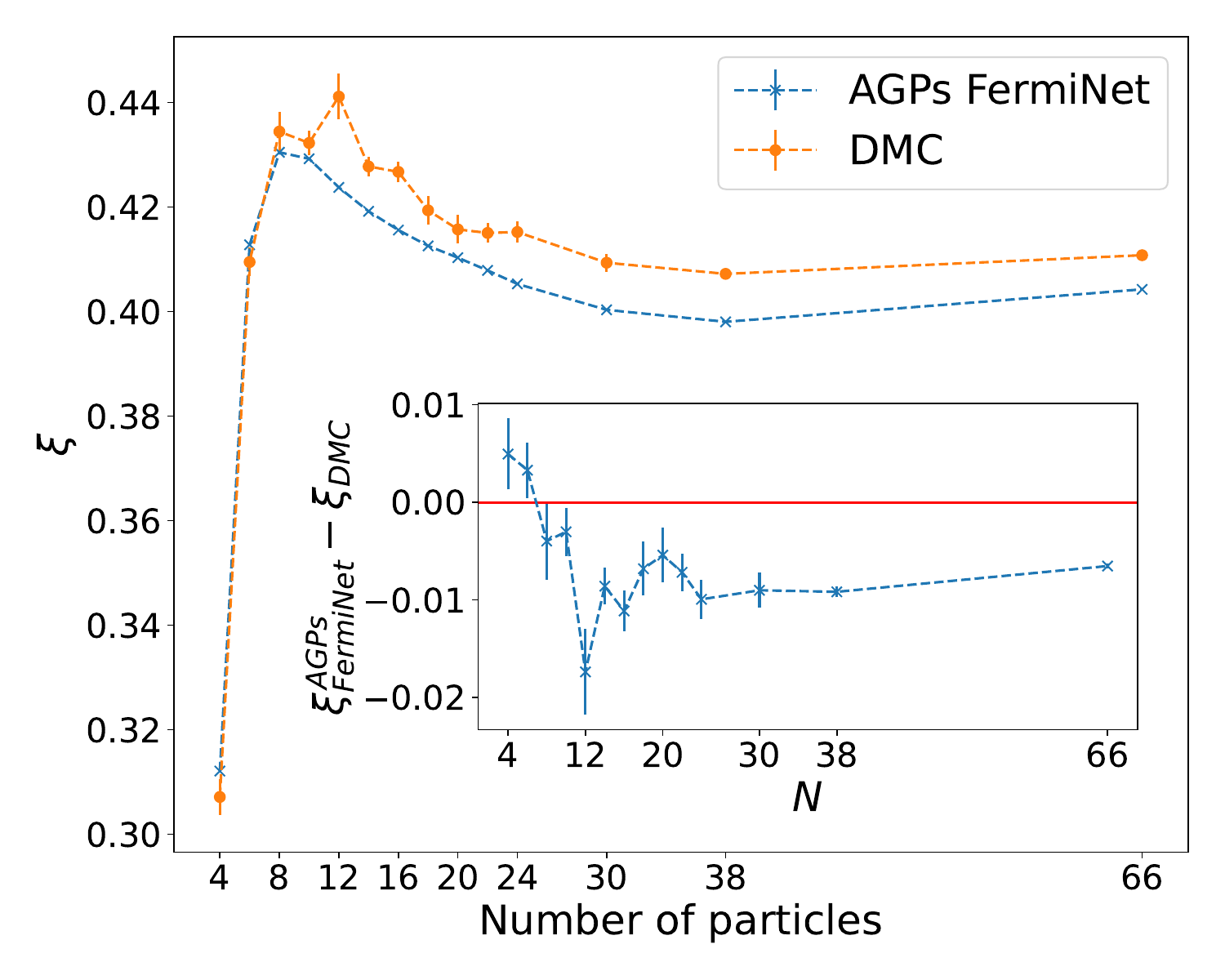}
    \caption{Comparison of the system-size dependent values of the Bertsch parameter, $\xi$, as calculated using the AGPs FermiNet and FN-DMC, with $k_F=1$ and $\mu = 12$. According to the variational principle, lower values are better.
    The error bars on the AGPs FermiNet results are smaller than the sizes of the crosses.
    Inset: difference between the AGPs and FN-DMC values of the Bertsch parameter.
    The errors in the inset are obtained by adding the standard errors of the FN-DMC and AGPs FermiNet results in quadrature. The latter are obtained by computing the standard error of the MCMC-averaged Bertsch parameter accumulated over 50,000 inference steps.
    }
    \label{fig:xi_dmc}
\end{figure}

We next compare our results with the state-of-the-art FN-DMC results of Forbes \textit{et al.}~\cite{forbes2011}, shown in Fig.~\ref{fig:xi_dmc} for the case $k_F = 1$ and $\mu=12$.
The AGPs FermiNet achieves a lower energy per particle than FN-DMC for all system sizes except for $N=4$ and $N=6$. The dependence of the Bertsch parameter on system size is also smoother when calculated with the AGPs FermiNet
\footnote{
  Limited by the scope of this work, we did not proceed to carry out a similar set of calculations done by Forbes et al. \cite{forbes2011} for extrapolating the results to $\mu \rightarrow \infty$.
}.
A full training curve of $N=66$ with comparison to the FN-DMC energy can be found in Appendix~\ref{appendix:training_curve_66ufg}.

\begin{figure}[ht!]
    \centering
    \includegraphics[width=0.9\columnwidth]{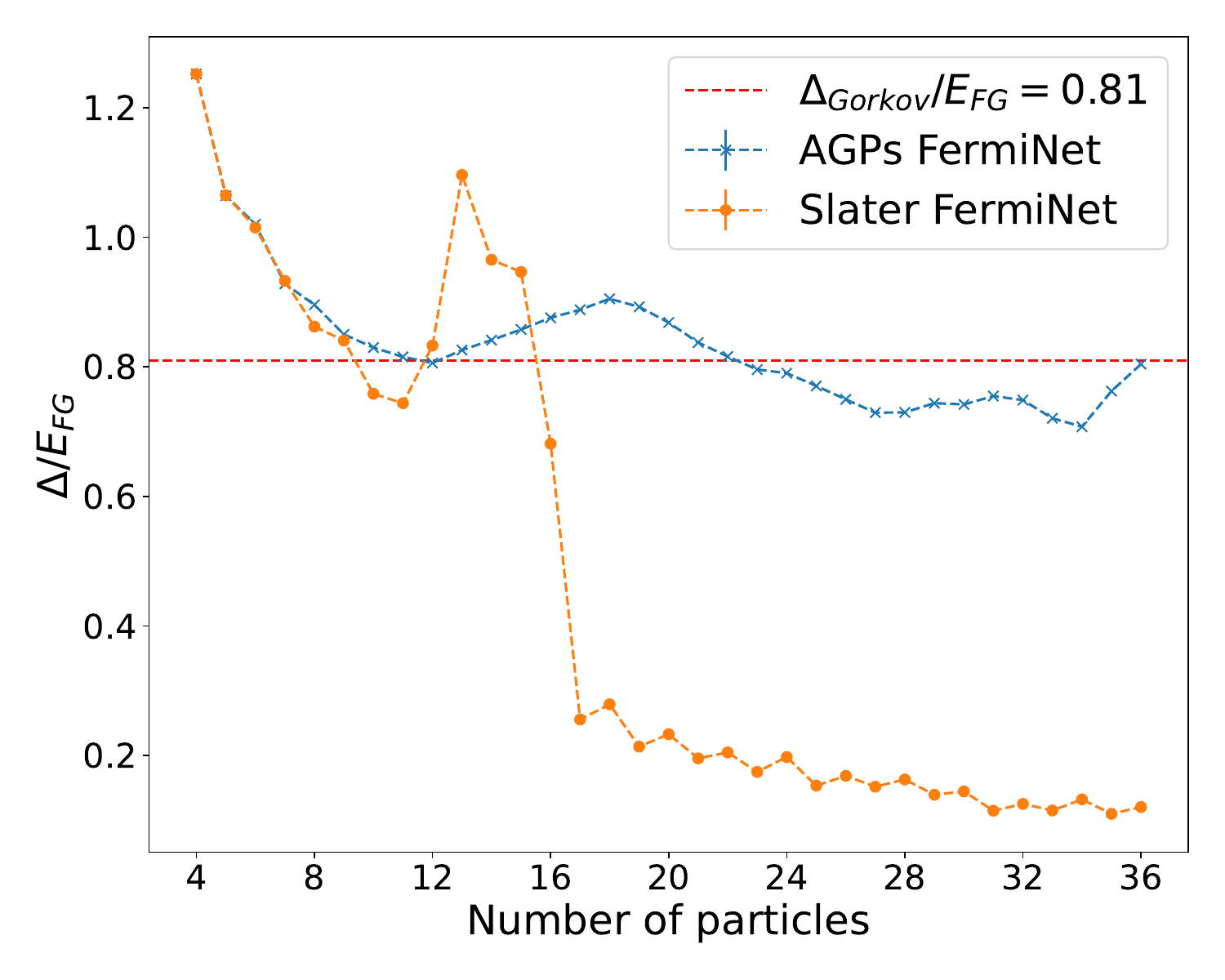}
    \caption{Pairing gaps calculated with the Slater FermiNet and the AGPs FermiNet for different numbers of particles $N$, with $r_s=1$ and $\mu = 12$.}
    \label{fig:pairing_gap}
\end{figure}

The pairing gap may be found using the approximation formula \cite{palkanoglou2020,carlson_review_2013}
\begin{equation}
    \Delta = (-1)^N\left [E(N+1) - \frac{1}{2}\left [E(N) + E(N+2) \right ] \right ],
\end{equation}
where $N$ is the total number of particles in the box.
The results from $N=4$ to $N=36$ are shown in Fig.~\ref{fig:pairing_gap}. 
Also shown is the thermodynamic ($N \rightarrow \infty$) limit of the BCS pairing gap including Gorkov's polarization correction \cite{gorkov1961}:
\begin{align}
    \Delta_\text{BCS} &= \frac{8}{e^2} \frac{\hbar^2 k_F^2}{2m}\exp(\frac{\pi}{2k_F a}) , \\
    \Delta_\text{Gorkov} &= \frac{1}{(4e)^{1/3}} \Delta_\text{BCS} .
\end{align}
Here $a$ is the scattering length of the interaction, which is infinite in the UFG.
In this limit, $\Delta_\text{BCS} = 1.804 E_\text{FG}$ and $\Delta_\text{Gorkov} = 0.815 E_\text{FG}$, where $E_\text{FG} = \frac{3}{5} \frac{\hbar^2 k_F^2}{2m}$ is the average energy per particle of an unpolarized non-interacting Fermi gas and $e$ is Euler's number
\footnote{
  Although writing the pairing gap in units of the Fermi energy would be more suitable as it is a one-body quantity, we choose to express it in units of $E_{FG}$ instead, because i) we want to be consistent with the units used in the paper by J.~Carlson \emph{et al.} \cite{carlson2003}, which also used $E_{FG}$ as the energy unit for the pairing gap, and ii) to keep the units consistent with other quantities in this work.
}.
The UFG is a strongly coupled system, so the BCS and Gorkov estimates of the gap need not be accurate.

The striking collapse of the pairing gap with increasing system size shows that the Slater FermiNet Ansatz struggles to describe paired states in systems of more than $10$ particles.
The AGPs FermiNet Ansatz behaves much better, although the oscillations with system size suggest that significant finite-size errors remain even for the largest systems simulated.

Another signature of fermionic superfluidity is the presence of off-diagonal long-ranged order in the two-body density matrix (TBDM), 
\begin{equation}
    \rho_{\uparrow\downarrow}^{(2)}(\mathbf{r}_1, \mathbf{r}_2; \mathbf{r}'_1, \mathbf{r}'_2)
    = \langle \hat{\psi}^\dagger_\uparrow(\mathbf{r}_1) \hat{\psi}^\dagger_\downarrow(\mathbf{r}_2) \hat{\psi}^{\phantom{\dagger}}_\downarrow(\mathbf{r}'_2) \hat{\psi}^{\phantom{\dagger}}_\uparrow(\mathbf{r}_1') \rangle ,
    \label{eq:tbdm}
\end{equation}
the largest eigenvalue of which diverges as the number of particles $N$ tends to infinity \cite{yang1962}.
The superfluid condensate fraction $c$ may be obtained by evaluating \cite{casino}
\begin{equation}
  c = \lim_{r\rightarrow\infty} \frac{\Omega^2}{N_\uparrow} \rho_{\uparrow\downarrow}^{(2)TR}(r) ,
    \label{eq:hr} 
\end{equation}
where $\Omega$ is the volume of the simulation cell, $N_\uparrow$ is the number of spin-up particles, and $\rho_{\uparrow\downarrow}^{(2)TR}(r)$ is the rotational and translational average of the TBDM
\begin{multline}
  \rho_{\uparrow\downarrow}^{(2)TR}(r) \\
  = \frac{1}{4\pi r^2 \Omega^2} \int\rho_{\uparrow\downarrow}^{(2)}(\mathbf{r}_1, \mathbf{r}_2; \mathbf{r}_1 + \mathbf{r}', \mathbf{r}_2 + \mathbf{r}') \delta(|\mathbf{r}'| - r) d\mathbf{r}_1 d\mathbf{r}_2 d\mathbf{r}' .
  \label{eq:avg_tbdm}
\end{multline}
The one-body density matrix, by contrast, tends to zero in the $r\rightarrow \infty$ limit \cite{yang1962}.
A full discussion of the methods used to evaluate the condensate fraction in QMC simulations can be found in the Appendix and the CASINO manual \cite{casino}.

\begin{figure}[ht!]
    \centering
    \includegraphics[width=0.9\columnwidth]{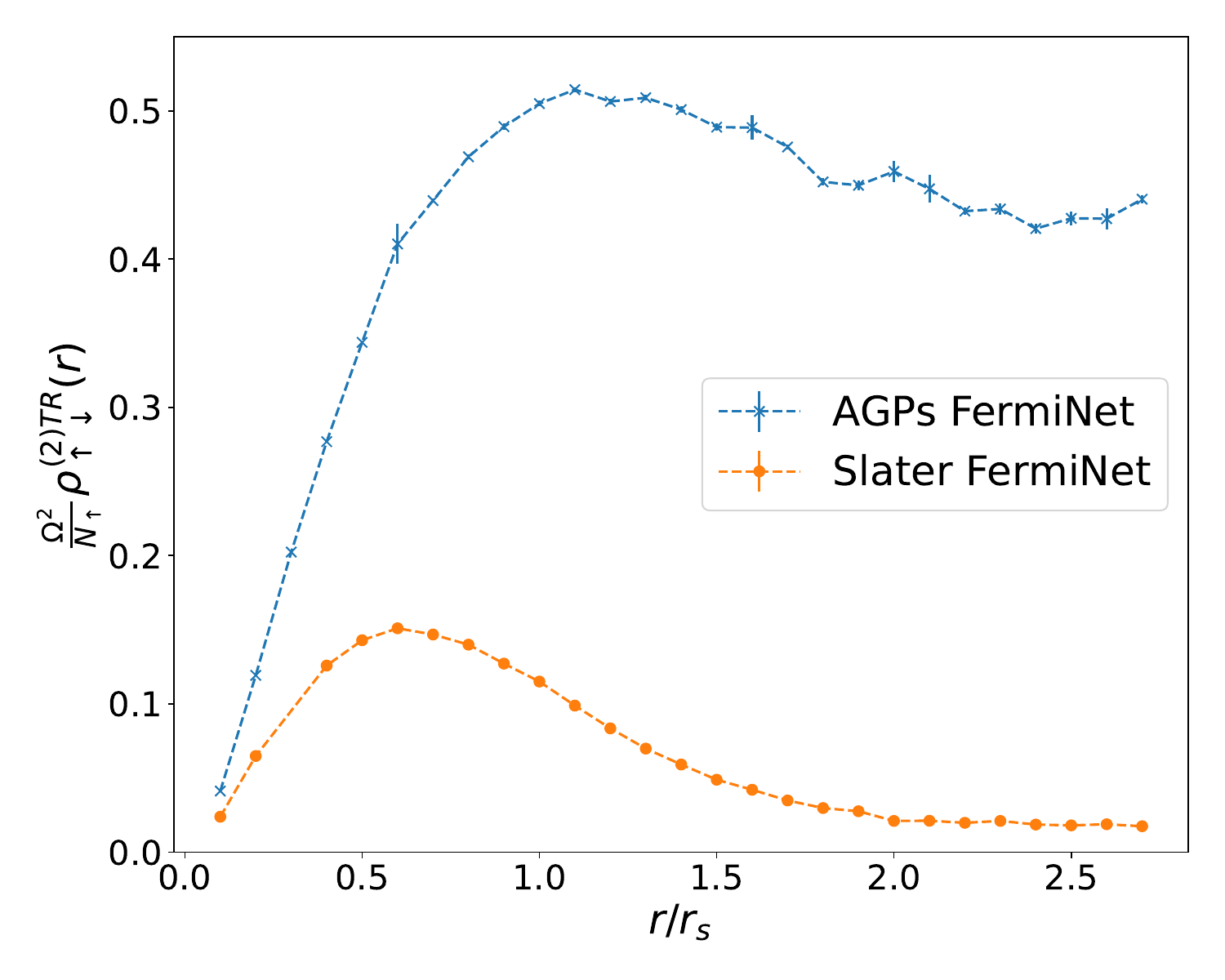}
    \caption{Comparison of the TBDM estimators calculated using the AGPs FermiNet and the Slater FermiNet with $N=38$, $r_s=1$ and $\mu = 12$. The error bars show the standard error of the TBDM estimator, accumulated over 2,000 inference steps. Most of the error bars are so small that they are obscured the symbols.}
    \label{fig:cfraction}
\end{figure}

After fully training both the Slater FermiNet and the AGPs FermiNet for the $N=38$ particle system, we used the resulting neural wave functions to compute the quantity $\frac{\Omega^2}{N_{\uparrow}} \rho_{\uparrow\downarrow}^{(2)TR}(r)$.
The results are shown in Fig.~\ref{fig:cfraction}, which provide further evidence that the Slater FermiNet fails to converge to the superfluid ground state; the quantity $\frac{\Omega^2}{N_{\uparrow}} \rho_{\uparrow\downarrow}^{(2)TR}(r)$ appears to be approaching zero in the large pair-separation limit, implying that the condensate fraction is also zero.
The same quantity for the AGPs FermiNet approaches a finite value which we estimated to be roughly $c=0.44(1)$ using the eight data points with separations $r/r_s \ge 2.0$.
This value is consistent with previous estimations from experiments and the most recent AFMC value from \cite{he_superfluid_2020} (Table~\ref{tab:cfraction_values}).

In addition, we also estimated the condensate fractions for the $N=66$ UFG at a fixed $\mu=12$ with two different densities: $r_s=1$ ($k_F r_e = 0.32$) and $k_F = 1$ ($k_F r_e = 0.17$), respectively.
We compute the quantity $\frac{\Omega^2}{N_{\uparrow}} \rho_{\uparrow\downarrow}^{(2)TR}(r)$ at five sequentially-spaced separations $r$ near $r=L/2$, where the quantity has approached its asymptotic value.
We then take the average of the five data points to get estimated values of the condensate fraction.
\footnote{
  Due to limited resources, we only compute five data points at far separations instead of the whole range of separations from $r=0$ to $r=L/2$ for $N=66$.
}.
Our estimate of the condensate fraction for $N=66$ is $c=0.42(1)$ at $k_F r_e=0.32$, and $c=0.52(1)$ at $k_F r_e=0.17$, which are both consistent with the experiments.
The results are summarized in Table~\ref{tab:cfraction_values}.
\begin{table*}[ht]
    \centering
    \begin{tabularx}{0.61\textwidth}{ l c }
        \midrule \midrule
        Method & Value \\
        \midrule
        Our estimate for $N=38$ at $k_F r_e=0.32$ & 0.44(1) \\
        Our estimate for $N=66$ at $k_F r_e=0.32$ & 0.42(1) \\
        Our estimate for $N=66$ at $k_F r_e=0.17$ & 0.52(1) \\
        FN-DMC for $N=38$ at $k_F r_e=0.03$ \cite{astrakharchik2005} & 0.61(2) \\
        FN-DMC for $N=66$ at $k_F r_e=0.03$ \cite{astrakharchik2005} & 0.57(2) \\
        FN-DMC for $N=128$ with VMC extrapolation at $k_F r_e=0.32$ \cite{morris_ultracold_2010} & 0.51\phantom{(1)} \\
        FN-DMC with $k_F r_e \rightarrow 0$ extrapolation for $N=66$ \cite{li2011} & 0.56(1) \\
        AFMC with $k_F r_e \rightarrow 0$ extrapolation for $N=66$ \cite{he_superfluid_2020} & 0.43(2) \\
        Experiment \cite{zwierlein2005} & 0.46(7) \\
        Experiment \cite{kwon2020} & 0.47(7) \\
        \midrule
    \end{tabularx}
    \caption{Estimates of the superfluid condensate fraction at unitarity using various methods. The quantity $k_F r_e$ is a dimensionless number, indicating the deviation of the simulated system from a perfect UFG with zero-range interaction.}
    \label{tab:cfraction_values}
\end{table*}

Although VMC methods are generally considered to be less accurate than FN-DMC methods, an important advantage of VMC methods is that almost any expectation value, including any reduced density matrix, may be estimated without bias.
The same is not true of FN-DMC simulations, which sample the wave function instead of its square modulus and produce biased ``one-sided'' estimates of the expectation values of operators that do not commute with the Hamiltonian \cite{foulkes2001}.
Thus, there are very few unbiased and accurate first-principles calculations of the condensate fraction.
Our approach, having both the advantages of VMC and surpassing the accuracy of DMC, provides solutions to these problems and a more accurate way to estimate general expectation values.

Finally, we study how the number of block-diagonal determinants required to achieve a given accuracy scales with the number of particles in the system.
We choose six even-particle systems from $N=4$ to $N=14$ and compare the energies obtained using linear combinations of multiple block-diagonal FermiNet Slater determinants against energies obtained using a \emph{single-determinant} (and thus single-geminal) FermiNet AGPs wave function.
All other hyperparameters are as given in Table~\ref{tab:hyperparams} (Appendix~\ref{appendix:ferminet}).
\begin{figure}[ht!]
    \centering
    \includegraphics[width=0.9\columnwidth]{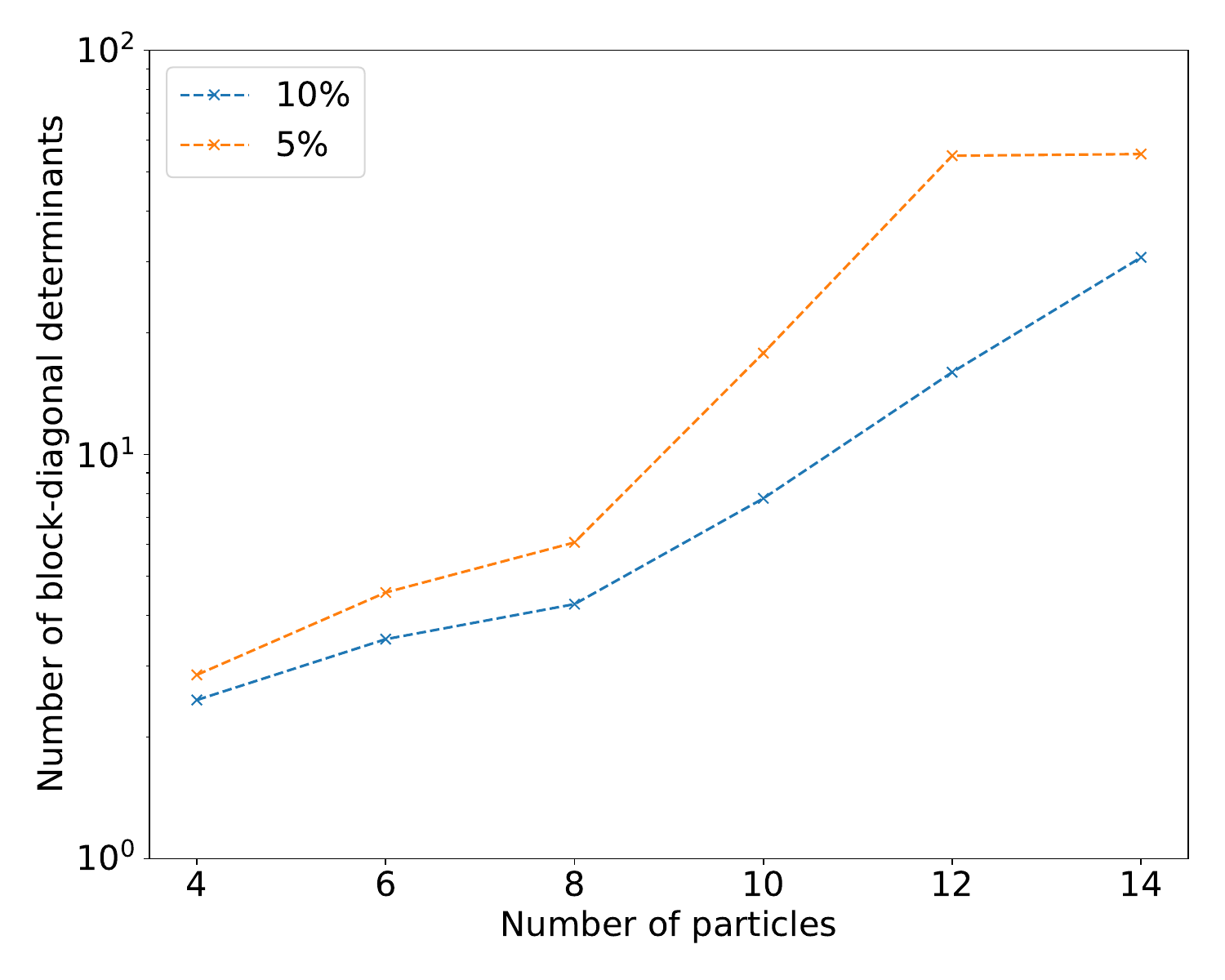}
    \caption{Number of block-diagonal determinants required for the multi-determinant Slater FermiNet energy to fall within 5\% and 10\% of the energy obtained using a single-determinant/single-geminal AGPs FermiNet.}
    \label{fig:ndet_nelec}
\end{figure}
In Fig.~\ref{fig:ndet_nelec}, we show that the number of block-diagonal determinants required to achieve a given percentage accuracy increases approximately exponentially with the number of particles.
Plots for each individual system, along with a more detailed discussion, can be found in Appendix~\ref{appendix:det-convergence}.
These results suggest that multi-determinant Slater FermiNet wave functions constructed using a neural network of fixed size are incapable of describing the ground state of the UFG accurately unless the number of block-diagonal determinants rises exponentially with system size. Hence, in practice, the AGPs FermiNet is required for studying paired systems.

\section{Discussion}
\label{section:discussion}

In this work, we used neural wave functions to study the benchmark superfluid system known as the UFG 
\footnote{
  Shortly after submitting this manuscript to the arXiv server, a closely related preprint on neural networks and the Fermi gas appeared \cite{kim2023}, proposing a similar extension to continuous-space neural network Ans{\"{a}}tzes.
}.
We showed that the Slater FermiNet Ansatz has difficulties in describing paired systems with strong, short-ranged attractive interactions between particles of opposite spin.
Hence, we proposed a way to improve the variational Ansatz by using determinants of FermiNet geminals, similar to an AGPs or a BCS wave function.
We showed mathematically that the Slater FermiNet is a limiting case of the AGPs FermiNet despite the use of fewer parameters in the latter. It follows that any FermiNet wave function can in principle be written as an AGPs FermiNet wave function.

We compared the total energies and energies per particle of the UFG as calculated using the Slater FermiNet and the AGPs FermiNet.
The former fails to produce a paired state when the number of particles, $N$, is greater than around $10$, while the AGPs FermiNet works very well.

As the UFG has a superfluid ground-state, we computed the pairing gap and condensate fraction for the $N=38$ system and compared estimates made with the Slater FermiNet and the AGPs FermiNet.
There is a clear qualitative difference between the pairing gap obtained using the AGPs FermiNet and the Slater FermiNet, with the latter approaching zero as the number of particles $N$ increases.
Calculations of the superfluid condensate fraction show a similar behavior: the AGPs FermiNet gives an accurate finite result, while the value obtained using the Slater FermiNet tends to zero in the limit of large system size.
Although the AGPs pairing gap shows significant finite size errors, it lies close to the mean-field BCS result with Gorkov-Melik-Barkhudarov corrections \cite{gorkov1961}.
Taken together, these results show that the Slater FermiNet is unable to represent large systems with superfluid ground states.
The AGPs FermiNet is much more suitable for studying paired systems such as the UFG.

To demonstrate the success of the AGPs FermiNet, we also compared our calculated total energies with state-of-the-art fixed-node diffusion QMC energies obtained using a Jastrow-BCS Ansatz \cite{forbes2011}.
For all systems with more than a few particles, the AGPs FermiNet achieves lower (\emph{i.e.}, better) variational energies than FN-DMC using the same model interaction and system parameters.

The inability of the Slater FermiNet Ansatz to accurately describe the UFG ground state comes as a surprise because the original FermiNet paper \cite{pfau2020} argued that any many-body fermionic wave function can be represented as a single determinant of FermiNet orbitals.
However, the mathematical argument relies on the construction of FermiNet orbitals with unphysical discontinuities.
Whether or not any wave function can be represented as a single determinant of FermiNet orbitals of the type used in practice, which are differentiable everywhere except at electron-electron and electron-nuclear coalescence points, remains an open question.
Another limitation is that the architecture of the FermiNet neural network, which is rather simple, may not be able to represent an arbitrary many-electron FermiNet orbital.
Even if a single-determinant Slater FermiNet wave function is general in principle, there is no guarantee that it is equally easy to represent all wave functions.
It may be that producing an accurate representation of a paired wave function requires the width and number of layers in the neural network to increase rapidly with system size.
Furthermore, if a network of fixed size is used, it may be necessary to increase the number of Slater FermiNet determinants rapidly as the system size increases.
The observation that the Slater FermiNet works well when $N \lessapprox 10$ but that the quality of the results degrades rapidly for larger systems, along with the scaling study presented in the final part of the Results section, suggests that this is, in fact, the case.

The AGPs FermiNet introduced in this paper shares many of the strengths of the Slater FermiNet.
In particular, there is no need to construct and optimize a new basis set for every new system or particle type.
If the AGPs FermiNet proves equally successful in other paired systems, it may now be relatively easy to investigate the importance of pairing in molecules, electron-positron systems, electron-hole liquids, and other $s$-wave superfluids.
Another strength of the AGPs FermiNet is the ease with which it is possible to optimize linear combinations of determinants of FermiNet pairing orbitals, such as the one in Eq.~(\ref{eq:agpsferminet_uppaired}).
This is much more difficult to accomplish with conventional wave functions based on explicit two-electron pairing orbitals or pairing orbitals represented as outer products of single-particle orbitals or basis functions.
Just as the many-particle orbitals in a Slater FermiNet radically generalize single-particle orbitals by incorporating electron-electron terms in a permutation-equivariant fashion, so the pairing functions in an AGPs FermiNet generalize BCS-style pairing functions by incorporating the effects of the remaining electrons in a permutation-equivariant fashion.

The AGPs FermiNet introduced here has a straightforward Pfaffian extension and can thus be applied to non-$s$-wave and triplet pairing.
Therefore, we expect it to become a powerful tool for understanding strongly correlated non-$s$-wave superfluid and superconducting systems such as Helium-3 or high-$T_c$ and $p$-wave superconductors.
Finally, our approach is not limited to the FermiNet neural network and can be readily adapted to use more recent architectures such as the Psiformer \cite{vonglehn2023}, GLOBE and MOON \cite{gao2023}, and DeepErwin \cite{scherbela2022}.

\section*{Acknowledgments}

We thank Stefano Gandolfi and Michael M.\ Forbes for providing the diffusion Monte Carlo data.
We gratefully acknowledge the European Union's PRACE program for the award of computing resources on the \href{https://apps.fz-juelich.de/jsc/hps/juwels/booster-overview.html}{JUWELS Booster supercomputer} in J\"{u}lich; the \href{https://www.hpc-rivr.si}{HPC RIVR consortium} and \href{https://eurohpc-ju.europa.eu}{EuroHPC JU} for resources on the Vega high performance computing system at \href{https://www.izum.si}{IZUM}, the Institute of Information Science in Maribor; and the UK \href{https://www.ukri.org/councils/epsrc/}{Engineering and Physical Sciences Research Council} for resources on the \href{https://www.baskerville.ac.uk}{Baskerville Tier 2 HPC service}.
Baskerville was funded by the EPSRC and UKRI through the World Class Labs scheme (EP/T022221/1) and the Digital Research Infrastructure programme (EP/W032244/1) and is operated by Advanced Research Computing at the University of Birmingham. 
We also gratefully acknowledge the \href{www.gauss-centre.eu}{Gauss Centre for Supercomputing e.V.} for funding this project by providing computing time through the John von Neumann Institute for Computing (NIC) on the GCS Supercomputer JUWELS at J\"{u}lich Supercomputing Centre (JSC).
JK is part of the Munich Quantum Valley, which is supported by the Bavarian state government with funds from the Hightech Agenda Bayern Plus.
WTL is supported by an Imperial College President's PhD Scholarship; 
HS is supported by the Aker Scholarship; and GC is supported by the UK Engineering
and Physical Sciences Research Council (EP/T51780X/1).
We also acknowledge the support of the Imperial-TUM flagship partnership.

\bibliographystyle{apsrev4-2}
\bibliography{main}

\clearpage
\newpage
\begin{appendix}

\section{Antisymmetric Geminal Power Wave Function}

This section discusses the relation between the antisymmetric geminal power singlet wave function (AGPs) and the BCS wave function.

\subsection{Fixed-particle Number BCS Ground State} \label{appendix:fixed_n_bcs}

The antisymmetrized geminal power single (AGPs) wave function may be obtained by projecting the BCS ground-state wave function \cite{bcs} into a subspace of fixed particle number \cite{bouchaud_pair_1988}.
We start with the BCS ground state:
\begin{align*}
  \ket{\Psi_{\text{BCS}}} 
  &= \prod_\mathbf{k} (u_\mathbf{k}^{\phantom{\dagger}} + v_\mathbf{k}^{\phantom{\dagger}} 
    \hat{c}^\dagger_{\mathbf{k}, \uparrow } 
    \hat{c}^\dagger_{-\mathbf{k},\downarrow})\ket{0} \\
  &= \bigg(\prod_\mathbf{k} u_\mathbf{k}^{\phantom{\dagger}} \bigg) 
    e^{
      \sum_\mathbf{k} \varphi_\mathbf{k}^{\phantom{\dagger}} 
      \hat{c}^\dagger_{\mathbf{k},\uparrow} 
      \hat{c}^\dagger_{-\mathbf{k},\downarrow}
    } \ket{0} ,
\end{align*}
where $\varphi_\mathbf{k} = \frac{v_\mathbf{k}}{u_\mathbf{k}}$.
Ignoring global phase factors and coefficients, the fixed-particle BCS wave function can be written as:
\begin{equation}
  \ket{\Psi_{\text{PBCS}}} 
  = \bigg(
    \sum_\mathbf{k} \varphi_\mathbf{k}^{\phantom{\dagger}} 
      \hat{c}^\dagger_{\mathbf{k},\uparrow} 
      \hat{c}^\dagger_{-\mathbf{k},\downarrow}
    \bigg)^p \ket{0} ,
\end{equation}
where $p = N/2$ is the number of pairs in the system and $N$ is the number of electrons.
After Fourier transforming this wave function, we get the real space wave function \cite{bouchaud_pair_1988}:
\begin{equation}
  \Psi_{\text{PBCS}} = \mathcal{A}[\Phi(1,1')\Phi(2,2')\cdots \Phi(p, p')] ,
  \label{eq:pbcs_anti}
\end{equation}
where $\mathcal{A}$ is the antisymmetrizer, the wave function corresponding to $\Phi(i,i')$ is
\begin{align}
  \Phi(i,i') &\equiv \Phi(\mathbf{r}_i, \sigma_i; \mathbf{r}_{i'} \sigma_{i'}) \notag \\
  &= \varphi(\mathbf{r}_i,\mathbf{r}_{i'}) \times \bra{\sigma_i\sigma_{i'}} \frac{1}{\sqrt{2}} (\ket{\uparrow\downarrow} - \ket{\downarrow\uparrow}) ,
\end{align}
and $\varphi(\mathbf{r}_i,\mathbf{r}_{i'}) = \sum_\mathbf{k} \varphi_\mathbf{k} e^{i\mathbf{k}\cdot(\mathbf{r}_i - \mathbf{r}_{i'})}$ is the Fourier transform of $\varphi_\mathbf{k}$.

As mentioned in the main text, Eq.~(\ref{eq:pbcs_anti}) can be written as a determinant of the pairing function $\varphi(\mathbf{r}_i,\mathbf{r}_{i'})$ after spin-assignment \cite{bouchaud_pair_1988,casula_geminal_2003}
\begin{equation}
  \Psi_{\text{PBCS}} = \det\left[\varphi(\mathbf{r}^\uparrow_i,\mathbf{r}^\downarrow_j)\right].
  \label{eq:pbcs_det}
\end{equation}

\section{Experimental Setup}

In this section, we report the FermiNet setup and hyperparameters used in this work.

\subsection{FermiNet}\label{appendix:ferminet}
The periodic version of the FermiNet implemented by Cassella \emph{et al.}~\cite{gcassella2022}, which can be found in the FermiNet repository \cite{ferminet_github}, was used as the basis for the AGPs code.
The small modifications required to support AGPs wave functions were made using the JAX Python library \cite{jax2018github}.
For the majority of our calculations, four NVIDIA A100 GPUs were used. For systems with $N>30$ particles, we used four nodes with a total of sixteen A100 GPUs to speed up the calculations.
A JAX implementation of the Kronecker-factored approximate curvature (KFAC) gradient descent algorithm \cite{pmlr-v37-martens15,kfac-jax2022github} was used for optimization.
The initial parameters of the network are initialized using Xavier (random) initialization\cite{glorot2010} and the positions of the particles are initialized uniformly in the simulation box.
We do not observe significant run-to-run variation in the final energy of a given system as a function of the random initialization.
The FermiNet hyperparameters are shown in Table (\ref{tab:hyperparams}) and the network sizes in Table (\ref{tab:network_hyperparameters}).
All training runs used $3 \times 10^5$ iterations to ensure convergence, except for the $N=66$ system (see Appendix~\ref{appendix:training_curve_66ufg}). When evaluating expectation values with an optimized wave function, $5 \times 10^4$ inference steps were used.
\begin{table}[ht]
  \centering
  \begin{tabularx}{0.45\textwidth}{ c c }
    \midrule \midrule
    Parameter & Value \\
    \midrule
    Batch size & 4096 \\
    Training iterations & 3e5 \\
    Pretraining iterations & None \\
    Learning rate & $(2e4 + t)^{-1}$ \\
    Local energy clipping  & 5.0 \\
    KFAC Momentum & 0 \\
    KFAC Covariance moving average decay & 0.95 \\
    KFAC Norm constraint & 1e-3 \\
    KFAC Damping & 1e-3 \\
    MCMC Proposal std.\ dev.\ (per dimension) & 0.02 \\
    MCMC Steps between parameter updates & 10 \\
    \midrule
  \end{tabularx}
  \caption{Hyperparameters used in all simulations.}
  \label{tab:hyperparams}
\end{table}

\begin{table}[ht]
  \centering
  \begin{tabularx}{0.42\textwidth}{ c c c }
    \midrule \midrule
    Parameter & Symbol & Value \\
    \midrule
    One-electron Stream Network Size & $n_l$ & 512 \\
    Two-electron Stream Network Size & N/A & 64 \\
    Number of Network Layers & $L$ & 4 \\
    Number of Determinants & $D$ & 32 \\
    \midrule
  \end{tabularx}
  \caption{Network sizes and number of determinants used in all simulations. The corresponding mathematical symbols mentioned in the main text of the paper, where available, are also listed.}
  \label{tab:network_hyperparameters}
\end{table}

\subsection{Estimation of the Two-Body Density Matrix}
The two-body density matrix (TBDM) in first quantized notation can be written as
\begin{multline}
  \rho_{\alpha\beta}^{(2)}(\mathbf{r}_1, \mathbf{r}_2; \mathbf{r}'_1, \mathbf{r}'_2) \\
  = N_\alpha (N_\beta - \delta_{\alpha\beta}) \frac{\int |\Psi(\mathbf{R})|^2 \frac{\Psi(\mathbf{r}'_1, \mathbf{r}'_2)}{\Psi(\mathbf{r}_1,\mathbf{r}_2)}d\mathbf{r}_3 \dots d\mathbf{r}_N}{\int |\Psi(\mathbf{R})|^2 d\mathbf{R}} ,
  \label{eq:tbdm_first}
\end{multline}
where $\alpha$ and $\beta$ denote the spin or particle species.
The superfluid condensate fraction in a finite and periodic system is defined as
\begin{equation}
  c = \frac{\Omega^2}{N_\alpha} \lim_{r\rightarrow \infty} \rho_{\alpha\beta}^{(2)TR}(r) ,
  \label{eq:cfraction_one}
\end{equation}
where $\Omega$ is the volume of the simulation cell, $N_\alpha$ is the number of particles with spin $\alpha$, and $\rho_{\alpha\beta}^{(2)TR}(r)$ is the translational and rotational average of the TBDM given in Eq.~(\ref{eq:avg_tbdm}).

The one-body density matrix (OBDM) is expected to tend to zero as $r \rightarrow \infty$.
However, because of finite-size effects, the OBDM is not necessarily zero within our simulation cell.
We therefore use an improved estimator in Eq.~(\ref{eq:modified_tbdm}) that removes the one-body contribution explicitly \cite{casino}:
\begin{widetext}
\begin{equation}
    \rho_{\alpha\beta}^{(2)}(\mathbf{r}_1, \mathbf{r}_2; \mathbf{r}'_1, \mathbf{r}'_2) = N_\alpha (N_\beta - \delta_{\alpha\beta}) \frac{\int |\Psi(\mathbf{R})|^2 
    \left[
        \frac{\Psi(\mathbf{r}'_1, \mathbf{r}'_2)}{\Psi(\mathbf{r}_1,\mathbf{r}_1)} - \frac{\Psi(\mathbf{r}'_1, \mathbf{r}_2)}{\Psi(\mathbf{r}_1,\mathbf{r}_2)} \frac{\Psi(\mathbf{r}_1, \mathbf{r}'_2)}{\Psi(\mathbf{r}_1,\mathbf{r}_2)}
    \right] d\mathbf{r}_3 \dots d\mathbf{r}_N}{\int |\Psi(\mathbf{R})|^2 d\mathbf{R}}.
    \label{eq:modified_tbdm}
\end{equation}
\end{widetext}
This quantity can then be estimated using Monte Carlo sampling.

\section{How many block-diagonal determinants does the Slater FermiNet need to achieve the same accuracy as the AGPs FermiNet with one determinant?}\label{appendix:det-convergence}

In the main text, we have demonstrated that the Slater FermiNet with 32 block-diagonal determinants is able to capture superfluidity in small systems but fails at larger systems.
Therefore, we are interested in the scaling of the original block-diagonal determinant FermiNet wave function with respect to the system size, \emph{i.e.} how many block-diagonal determinants do we need in order for the Slater FermiNet to converge to the ground state at each system size?

\begin{figure*}[ht!]
    \centering
    \includegraphics[width=2.0\columnwidth]{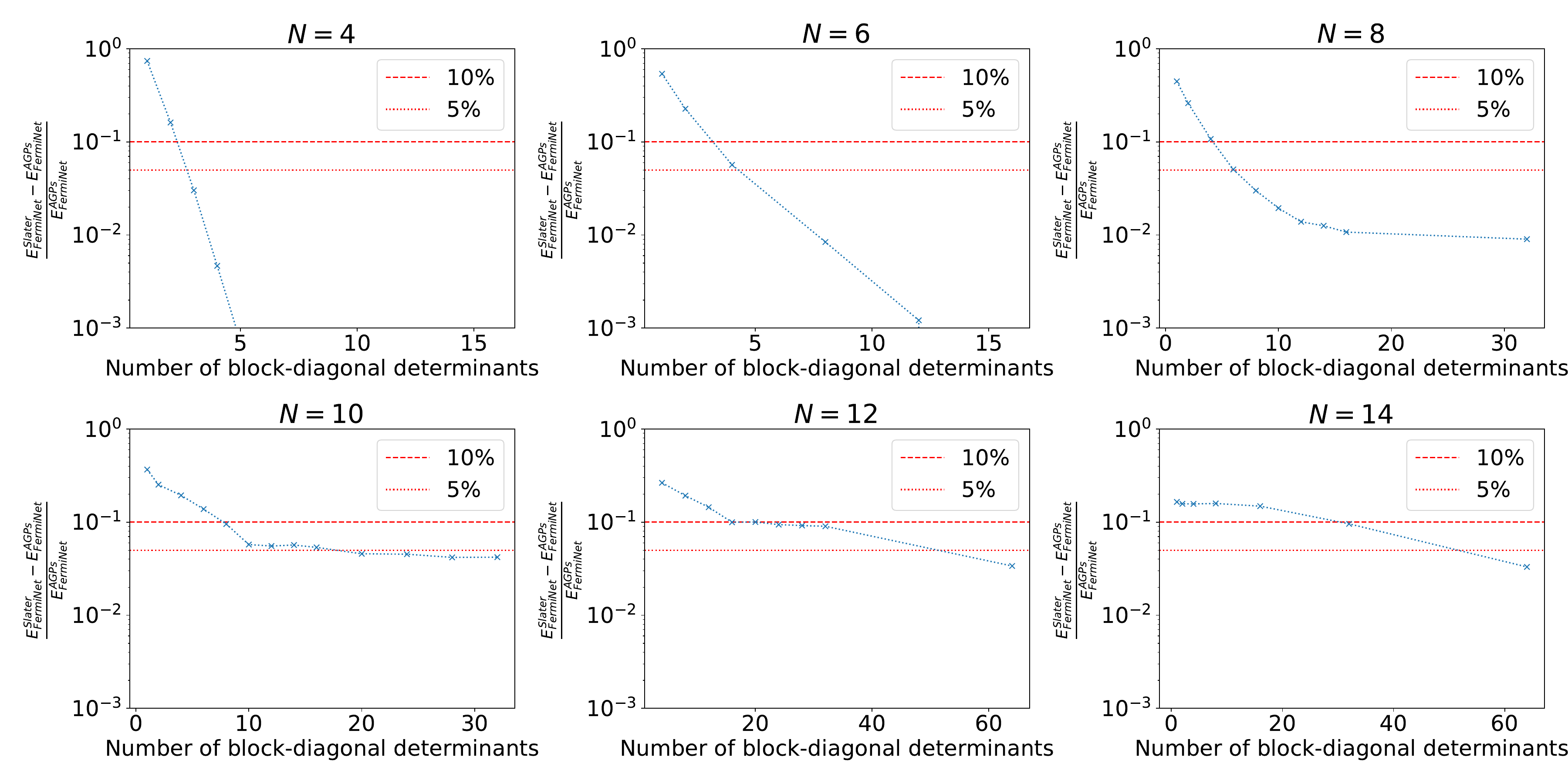}
    \caption{Percentage difference between the Slater FermiNet with various number of block-diagonal determinants and the AGPs FermiNet with one determinant at different system sizes, from 4 to 14 particles.}
    \label{fig:e_ndet_agps}
\end{figure*}

To answer this question, first we set the AGPs FermiNet with one determinant energies as baselines and plot the percentage difference of the Slater FermiNet from the baseline against the number of block-diagonal determinants used in the Slater FermiNet wave function, repeated at different system sizes from 4 to 14 even particles systems.
The results are shown in Fig.~\ref{fig:e_ndet_agps}.
 
As the results suggested, it becomes more difficult for the Slater FermiNet to get close to the AGPs FermiNet baseline as the number of determinant increases, especially for larger systems. This is due to the limited performance of the optimizer as the number of determinant increases.
Hence, due to the constraints on time and resources, it is not feasible to continue to increase the number of determinants until the Slater FermiNet achieve the same accuracy as the AGPs FermiNet. 
Instead, we decided to set two thresholds for the percentage difference between the two results. 
Here, we have used 5\% and 10\% as the thresholds, plotted as the two horizontal lines in Fig.~\ref{fig:e_ndet_agps}.
By plotting the x-intercept of the curves with the two threshold lines against system size, we can determine the relationship between the two, as shown in Fig.~\ref{fig:ndet_nelec}. 
As the y-axis in Fig.~\ref{fig:ndet_nelec} is in logarithmic scale, a roughly linear relationship suggests an exponential scaling of the number of block-diagonal determinants as system size increases.

The result indicates that, in theory, the Slater FermiNet is capable of converging to the ground state given the number of block-diagonal determinants is sufficiently large. 
In practice, the number of block-diagonal determinants required to learn the ground state increase exponentially as system size gets bigger and the number get inaccessible rapidly. 


\section{Dense determinant}
\label{appendix:dense}

Previous works \cite{gcassella2022,gerard2022,vonglehn2023,gao2022,ren2023,lin2023} have suggested that the use of dense determinants,
\begin{equation}
  \Psi^\text{Dense}_\text{FermiNet} = \det
    \begin{pmatrix}
      \phi_i^\uparrow(\mathbf{r}_j^\uparrow;\{\mathbf{r}_{/j}^\uparrow\};\{\mathbf{r}^\downarrow\}) 
      & \phi_i^\uparrow(\mathbf{r}_j^\downarrow;\{\mathbf{r}_{/j}^\downarrow\};\{\mathbf{r}^\uparrow\}) \\
      \phi_i^\downarrow(\mathbf{r}_j^\uparrow;\{\mathbf{r}_{/j}^\uparrow\};\{\mathbf{r}^\downarrow\}) 
      & \phi_i^\downarrow(\mathbf{r}_j^\downarrow;\{\mathbf{r}_{/j}^\downarrow\};\{\mathbf{r}^\uparrow\}) \\
    \end{pmatrix}
  \label{eq:dense-ferminet}
\end{equation}
in contrast to block-diagonal diagonal determinants with FermiNet in Eq.~\ref{eq:block-det-ferminet}, provide a slight gain in accuracy in various infinite and periodic systems.
In this subsection, we present a small set of calculations of the UFG with dense determinants and compare the results with the block-diagonal FermiNet and the AGPs FermiNet.
The results are presented in Table~\ref{tab:dense_det}.

\begin{table}[ht]
  \centering
  \begin{tabularx}{0.49\textwidth}{ c c c c c }
    \midrule \midrule
    Type of & \multicolumn{3}{c}{Total Energy [$E_{FG}$]} & Pairing \\
    Wave Function & $N=36$ & $N=37$ & $N=38$ & Gap [$E_{FG}$] \\
    \midrule
    Block & 17.8523(5) & 18.2032(5) & 18.3131(6) & 0.121(2) \\
    Dense & 17.7027(5) & 18.1601(5) & 18.2570(5) & 0.180(2) \\
    AGPs & 14.6059(4) & 15.8060(5) & 15.3975(5) & 0.804(1) \\
    \midrule
  \end{tabularx}
  \caption{Total energy of the UFG with 36 to 38 particles and their corresponding pairing gap using different wave functions.}
  \label{tab:dense_det}
\end{table}

Although the use of dense determinants in the UFG does provide lower energies comparing to the block-diagonal determinants, they are still significantly higher than the AGPs FermiNet energies. 
In addition, the qualitative behaviors, such as the absence of odd-even staggering, are still similar to the block-diagonal FermiNet, which are qualitatively different from the AGPs FermiNet.


\section{Translational symmetry of the AGPs FermiNet}
\label{appendix:translational_invariance}

In this section, we discuss the consequences of setting an origin in the simulation cell.

As mentioned in the main text, in all of our simulations, an origin is set at one corner of the simulation cell.
The particle coordinates $\mathbf{r}^\alpha_i$, which are used as inputs to the one-electron stream network, are taken with respect to this origin.
Thus, the AGPs FermiNet Ansatz does not impose translational invariance with respect to simultaneous translations of all particle coordinates.

As the model unitary Fermi gas system has a translationally invariant ground-state, it is natural to use a translationally invariant Ansatz to study it.
Whilst it is possible to neglect the one-electron stream network and only use the two-electron stream network, which is translationally invariant by construction, in the AGPs FermiNet simulations, we found that doing so gives significantly worse energies than the standard AGPs FermiNet.
This is due to the limitation on the expressivity of the two-electron stream network, which, despite being translationally invariant, is not sufficient to represent the ground-state of the unitary Fermi gas.
This finding is also consistent with the findings in the previous study of the homogeneous electron gas by G. Cassella, \emph{et al.} \cite{gcassella2022}.
Hence, we include the one-electron stream networks in our Ansatz when studying the unitary Fermi gas.

\begin{figure*}[ht!]
    \centering
    \includegraphics[width=\textwidth]{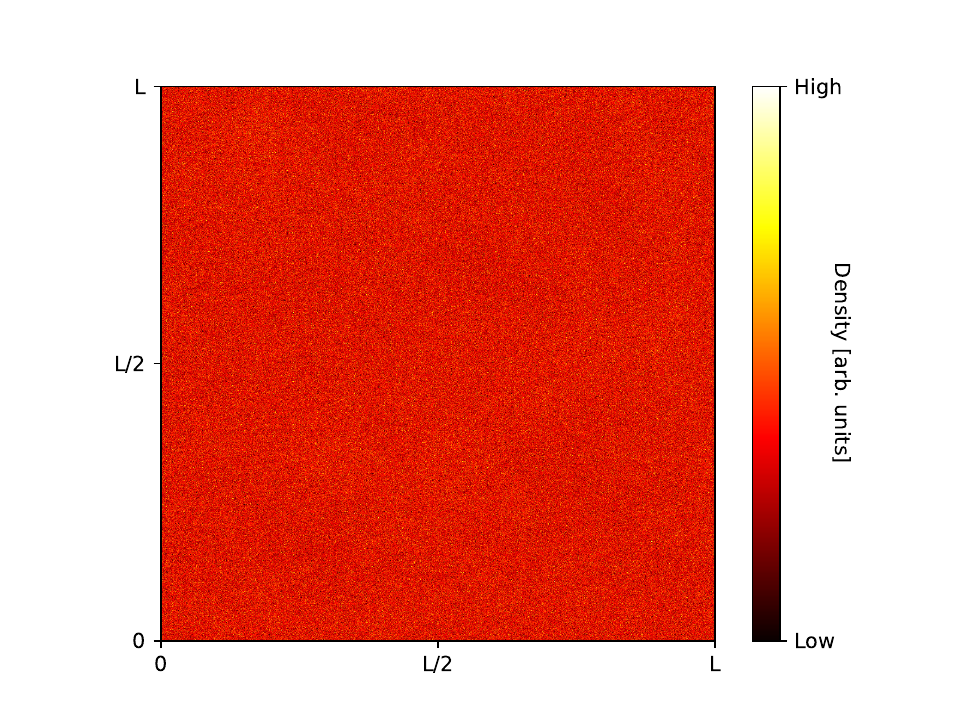}
    \caption{2D projection of the $N=14$ unitary Fermi gas density in the simulation cell from sampling the converged AGPs FermiNet wave function. $L$ is the length of the simulation cell.}
    \label{fig:density}
\end{figure*}

Despite an explicit origin being embedded into the one-electron stream network in the AGPs FermiNet Ansatz when studying the unitary Fermi gas, we have concluded that this does not create any obvious bias in the converged wave function.
Evidence supporting this is shown in Fig.~(\ref{fig:density}), where a 2D-projected density of the $N=14$ unitary Fermi gas, obtained using the converged AGPs FermiNet wave function, is plotted.
There is no obvious structure in the density plot, and is uniform across the whole simulation cell.
Hence, we believe the Ansatz is capable of representing translationally invariant wave functions, even though this is not explicitly imposed.


\section{Training curves of the 66 particle UFG}
\label{appendix:training_curve_66ufg}

We present the training curves of the $N=66$ UFG in this section.
Due to limited computational resources, we only trained the 66 particle UFG system up to 150,000 steps, which is half the number of steps we use to train all other systems in the paper (see Tab.~\ref{tab:hyperparams}).
Therefore, even though our $N=66$ results are able to outperform the FN-DMC results by a similar amount as the smaller systems, it is not fully converged as demonstrated in the training curve below. We emphasize that the energy we obtained is still a variational upper bound on the ground state energy.

\begin{figure}[ht!]
    \centering
    \includegraphics[width=0.9\columnwidth]{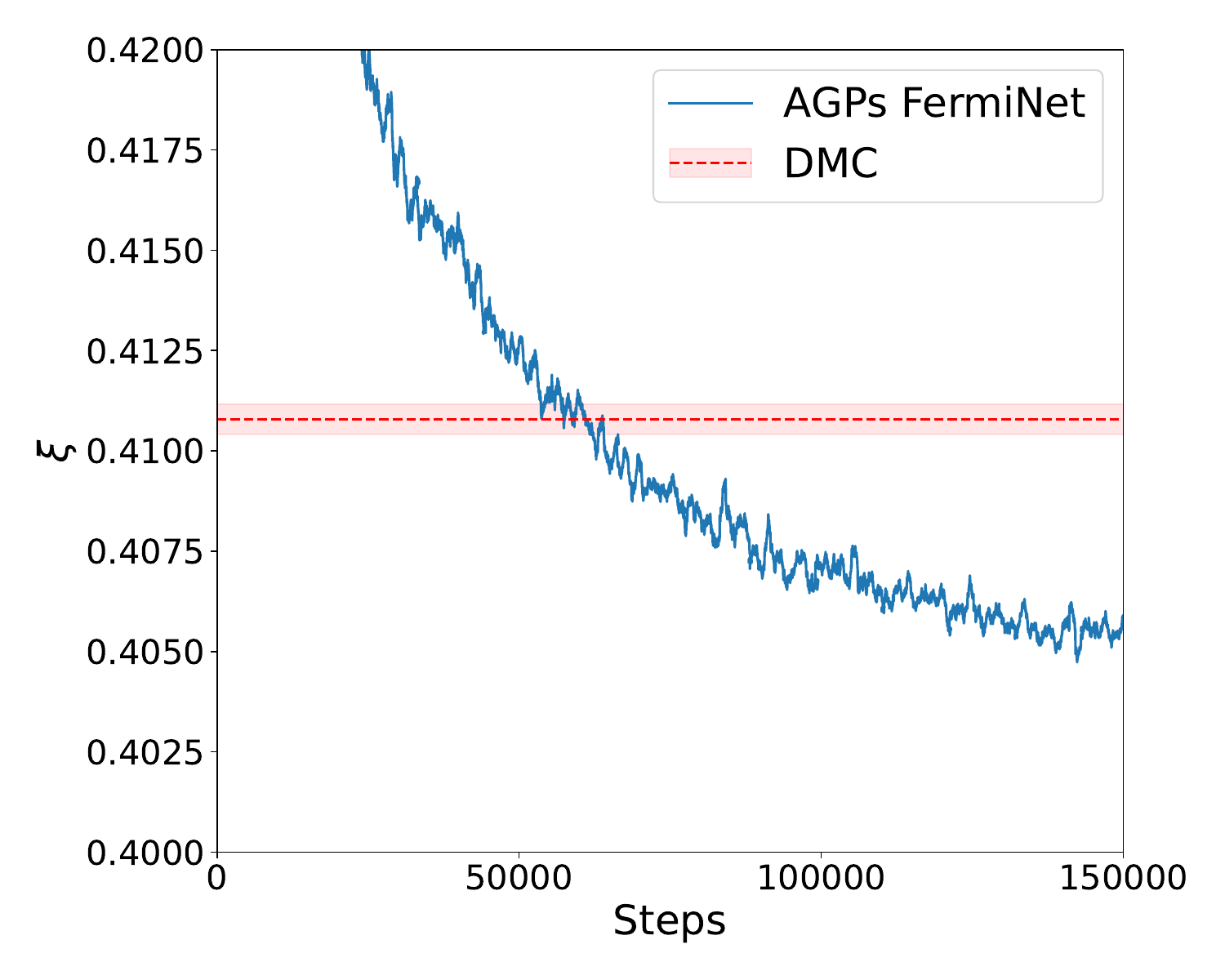}
    \caption{Training curve of the 66 particle UFG with 32 determinants and using the geminals in Eq.~\ref{eq:ggw2e}. Red dashed line is the FN-DMC result from \cite{forbes2011} under the same density and interaction width.}
    \label{fig:training_curve_66ufg}
\end{figure}

\end{appendix}

\end{document}